\newcommand\ha{{H$\alpha$}}
\newcommand\hb{{H$\beta$}}

\newcommand\kms{\:\rm{km\,s^{-1}}}

\newcommand\VEL{\:{\rm km\:s^{-1}}}

\newcommand\OiL{[\ion{O}{1}] $\lambda 6300$}

\newcommand\SiiLL{[\ion{S}{2}] $\lambda\lambda 6716, 6731$}
\newcommand\NiiLL{[\ion{N}{2}] $\lambda\lambda 6548, 6583$}

\newcommand\OiLL{[\ion{O}{1}] $\lambda\lambda 6300, 6363$}

\newcommand\sii{[\ion{S}{2}]}
\newcommand\nii{[\ion{N}{2}]}
\newcommand\oi{[\ion{O}{1}]}

\newcommand\oiii{[\ion{O}{3}]}
\newcommand\feii{[\ion{Fe}{2}]}
\newcommand\hii{\ion{H}{2}}

\documentclass[modern]{modaastex631}

\usepackage[utf8]{inputenc}
\usepackage{natbib}
\usepackage{comment}
\usepackage{todonotes}
\usepackage{graphicx}
\usepackage{lineno}
\turnoffediting

\begin{document}

\flushleft{Accepted for publication in {\em The Astrophysical Journal}}

\title{High-Resolution Spectra of Supernova Remnants in M83}

\correspondingauthor{P. Frank Winkler}
\email{winkler@middlebury.edu}

\author[0000-0001-6311-277X]{P. Frank Winkler}
\affil{Department of Physics, Middlebury College, Middlebury, VT, 05753; 
winkler@middlebury.edu}

\author[0000-0002-4134-864X]{Knox S. Long}
\affil{Space Telescope Science Institute,
3700 San Martin Drive,
Baltimore MD 21218, USA; long@stsci.edu}
\affil{Eureka Scientific, Inc.
2452 Delmer Street, Suite 100,
Oakland, CA 94602-3017}

\author[0000-0003-2379-6518]{William P. Blair}
\affil{The William H. Miller III Department of Physics and Astronomy, 
Johns Hopkins University, 3400 N. Charles Street, Baltimore, MD, 21218; 
wblair@jhu.edu}

\author[0000-0002-4596-1337]{Sean D. Points}
\affiliation{Cerro Tololo Inter-American Observatory, NSF's NOIRLab, Casilla 603, La Serena, Chile; 
sean.points@noirlab.edu}

\begin{abstract}
In order to better characterize the rich supernova remnant (SNR) population of M83 \edit1{(NGC\,5236)}, we have obtained high-resolution ($\sim 85\VEL$) spectra of 119 of the SNRs and SNR candidates in M83 with Gemini/GMOS, as well as new 
\ spectra of the young SNRs B12-174a and SN1957d.
Most of the SNRs and SNR candidates have \sii:\ha\ ratios that exceed 0.4.  
Combining these results with earlier studies we have carried out with MUSE and at lower spectroscopic resolution with GMOS, we have confirmed a total of 238 emission nebulae to be SNRs on the basis of their \sii:\ha\ ratios,  about half of which have emission lines that show velocity broadening greater than 100$\VEL$, providing a kinematic confirmation that they are SNRs and not \hii\ regions. Looking at the entire sample, we find a strong correlation between velocity widths and the line ratios of \oi$\lambda$6300:\ha, \nii$\lambda$6584:\ha\ and \sii$\lambda\lambda$6716,6731:\ha.  The density-sensitive \sii$\lambda$6716:$\lambda$6731 line ratio is strongly correlated with SNR diameter, but not with the velocity width.  We discuss these results in the context of previously published shock models.
\end{abstract}


\section{Introduction}
Supernova remnants (SNRs) in the  grand-design spiral galaxy M83 \edit1{ (NGC\,5236, also known as the Southern Pinwheel) }have been the subject of several recent studies, and there are currently some 364 catalogued SNRs.\footnote{Unless we specifically state, all of the object counts mentioned in this paper, regarding for example, the number of SNR candidates with spectra with \sii:\ha\  ratios greater than 0.4, {\em exclude} the two young SNRs B12-174a and SN1957d, which have spectra that are very different from those of ``ordinary'' ISM-dominated SNRs.} and candidates 
\citep[][and references in these papers]{blair12,winkler17,williams19,long22}.\footnote{The ``current'' sample is discussed by \cite{long22}; it excludes a number of objects suggested as candidates based on high \oiii:\hb\ ratios, largely because subsequent observations have not produced additional evidence that they were SNRs.}  Historically, the most common means for identifying a nebula as a SNR has been through a ratio of \SiiLL:\ha\ emission $\gtrsim 0.4$, the physical basis being that shocks such as present in SNRs produce a long cooling ``tail" where S$^+$ is abundant, compared with photoionized nebulae, where S is more highly ionized and hence the \sii:\ha\ ratio is lower (typically \sii:\ha\ $\sim 0.1$).  In M83, some 211 objects  have previously been spectroscopically confirmed to have \sii$\lambda\lambda$6716,6731:\ha\ $\ge$ 0.4,  which is more than in any other galaxy to date \citep{long22}.

Recently, \citet{points19} have explored a complementary SNR diagnostic based on kinematics: the widths of the emission lines.  For SNRs the line widths should be comparable to the shock velocity, $\gtrsim 100\kms$ for all but the most evolved SNRs, compared with the narrow lines expected from photoionized regions, where bulk velocities are typically much smaller.    This technique has been used recently by \citet{Mcleod21} to confirm SNRs in NGC\,300, and by \citet{long22} to confirm SNRs in the central region of M83, Both of these papers are based on data from the Multi-Unit Spectroscopy Explorer (MUSE) on the VLT.

The over-arching theme of the series of papers on SNRs in M83, by ourselves and others, is to define as complete a sample of confirmed SNRs in M83 as possible, and to use this sample to explore the properties of SNRs in general: the progenitor populations from which they arise, and their evolution as remnants until they merge back into the interstellar medium.  M83 is the galaxy that has been the most studied since, as the host to at least six (and likely seven)  SNe that have exploded since 1900 \citep{blair15}, it should have a rich SNR population.  Furthermore, M83 is nearly face-on, where all the objects are at essentially the same, well-measured distance of 4.61 Mpc \citep{saha06}, and its contents should be minimally affected by absorption.

In this paper, we report spectra for 119 SNRs and candidates from throughout M83 obtained using the Gemini Multi-Object Spectrograph (GMOS) on the 8.1 m Gemini-S telescope on Cerro Pachon, in a configuration that gave roughly twice the spectroscopic resolution of our earlier GMOS spectra \citep{winkler17}, in order to enable us to more effectively use the kinematic criterion in addition to the \sii:\ha\ ratio to investigate SNRs and SNR candidates.  This total includes spectra of 24 objects in the outer galaxy that had no previously reported spectra  We also extract the spectra of 36 \hii\ regions for comparison, some targeted with slitlets and others extracted from along the SNR slitlets.  The paper is organized as follows: section 2 describes the observations and data processing, section 3 presents results that follow directly from the observations, and section 4 discusses these results more broadly, in the context of the earlier data and shock models.  Section 5 provides a summary of our findings.

\section{Observations and Data Reduction}

Our observations were carried out using GMOS on Gemini-S during the 2019A, 2020A, and 2021A semesters.  (This protracted duration resulted from equipment and weather issues in 2019 and the 2020 COVID-19 shutdown of observatory operations.)  We were explicitly interested in using line broadening as a diagnostic for SNRs, and had determined that a velocity resolution higher than $\sim 100\kms$ would be required \citep{points19}.  Early in the 2019A semester we carried out a series of daytime tests of different grating, order, and slit width combinations to determine the configuration that would be most effective.  These showed that of the choices with sufficient resolution, the B1200 grating in first order gave by far the best throughput, and that a slit width of 0\farcs 6 gave a resolution of $85\kms$, barely degraded from the minimum slit width of 0\farcs 5.  We thus adopted this combination for all our observations.

The spectral range covered $\sim 1600$ \AA, with the exact range depending on the location of each object slit in the dispersion direction on the mask plate, but in all cases the 6000 - 7000 \AA\ range was covered.  This ensured that measurements of \OiLL, \NiiLL, \ha, and \SiiLL\ would be obtained for all objects regardless of the slit position on the mask.
For all observations, the Hamamatsu detector was binned $2 \times 2$, except for the final mask (No.\ 5), which was erroneously not binned.  The $2 \times 2$ binned data had a dispersion of 0.52 \AA\ per (binned) pixel for a spectral resolution of about 1.85 \AA\ (FWHM, $85 \kms$ at \ha, measured for isolated night-sky lines) and a spatial resolution of 0\farcs 16  per binned pixel.

\subsection{Masks for Multi-Object Spectra}
Masks were designed based on GMOS-S images taken in 2011; see \cite{winkler17}, which used these same images to plan the GMOS low-resolution spectra reported there.  The slit length varied, but was $\geq 6\arcsec$ in all cases.
Observations with a single mask were carried out in 2019A, and a second was completed before operations were shut down in 2020A.  It was not until 2021A that we were finally able to complete our program by obtaining spectra with the final three masks.  Table 1 gives a journal of these observations.

% Fig. 1
\begin{figure}
\plotone{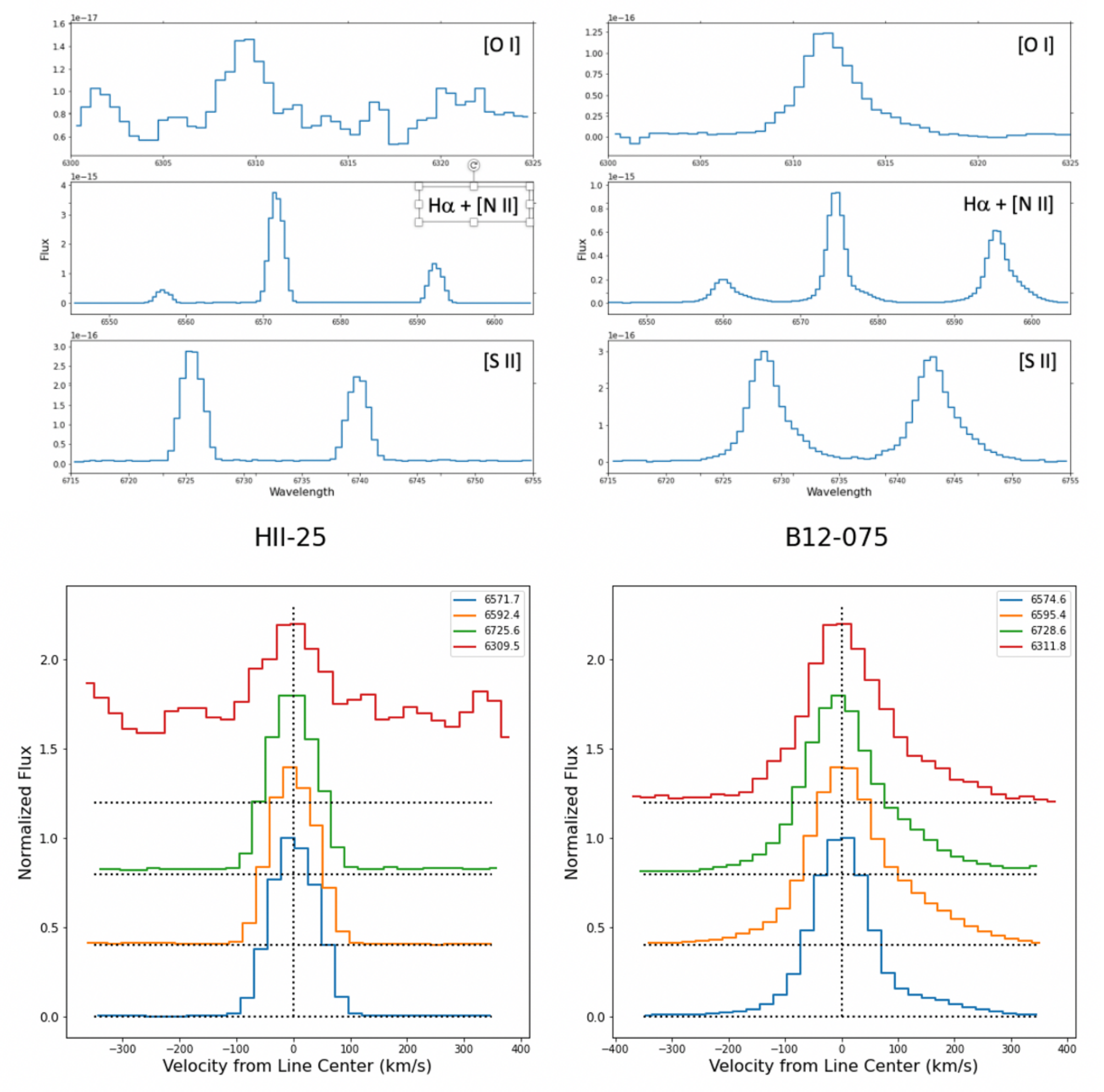}
\caption{GMOS spectral data for an \hii\ region (left) and a SNR (right) are compared in two different formats. Above, the regions around spectral lines of interest are plotted in separate panels as marked.  Below, representative spectral lines are shown normalized and stacked to better show their relative line profiles in velocity space.  The higher velocities for the SNR are especially visible in the wings of the lines. Note that the \oi\ line in the \hii\ region is exceedingly weak compared with the other lines.
\label{fig:example_spectra}}
\end{figure}

Each of our five masks targeted 20-25 SNRs or candidates.  For masks 1 and 2, observed in 2019 and 2020, many of the objects chosen were ones for which our earlier, lower-resolution spectra gave marginal \sii:\ha\ ratios, in order to investigate the utility of the velocity-width criterion in cases where \sii:\ha\ is not definitive.  For the final three masks observed in 2021, we concentrated on candidates which at the time of planning had no previous spectra, as well as candidates newly discovered with MUSE \citep [since published by][]{long22}.  
%In order to be able to compare spectra obtained of SNRs and SNR candidates, 
We also obtained spectra of several emission nebulae for which images did {\em not} indicate an enhanced \sii:\ha\ ratio for comparison.  A few of these are  relatively bright, compact \hii\ regions that we explicitly targeted with their own slitlets, but most are generally lower surface-brightness nebulae that appeared on portions of the slitlets  in close proximity to SNR candidates. In total, we observed 119 SNRs and candidates\footnote{Six objects were observed twice, for a total of 125 SNR spectra.} as well as 36 \hii\ regions; the positions and projected galactocentric distances of the latter group are listed in 
Table 2.

\subsection{Data Reduction}

The data were reduced using the standard {\tt gemini} package and methods in IRAF.\footnote{IRAF is distributed by the National Optical Astronomy Observatory, which is operated by the Association of Universities for Research in Astronomy, Inc., under cooperative agreement with the National Science Foundation.}  This included nightly bias frames, and quartz  flats and CuAr arc spectra, both taken immediately before or after each science observation.   Standard processing included separating the spectra from each slitlet, wavelength calibration, and combination of all the spectra for a given mask, which have staggered central wavelengths to cover inter-chip gaps.  We then examined the 2-D spectrum from each slitlet and subtracted the background sky, generally by fitting  a constant sky background corresponding to the minimum level as determined over critical wavelength regions.  Finally, we extracted 1-D spectra for each object by examining the 2-D spectra and summing over the spatial region corresponding to each object.

The resulting spectra are generally of excellent quality.  In Fig.~\ref{fig:example_spectra} we show a comparison of data for an \hii\ region and one of the SNRs, where the extra broadening in the SNR spectrum is obvious, extending out to several hundred $\VEL$.  
Fig.~\ref{fig:example_snrs} shows two additional SNRs with even broader line profiles and different amounts of asymmetry relative to the line peak. All three of the SNR spectra are very obviously broadened compared with the \hii\ region spectrum.  The details of these examples will be discussed below.

\section{Analysis and Results}

\subsection{Gaussian Fitting}

In order to characterize the spectra in a relatively uniform manner we have fit the the most prominent lines in the spectra with simple Gaussians.  Prior to fitting, we inspected each of the spectra and set a flag that excluded obviously discrepant points from the spectra.  Some of the spectra, especially ones from SNRs in or near the complex central region of M83, have some residual narrow \ha\ emission in addition to the broader \ha\ emission that is truly characteristic of the SNR itself.  In such cases, we attempted to exclude points associated with a narrow component of \ha, which had not been ``correctly'' subtracted as background.  We  fit \OiL\ as a singlet, but, as was the case when we analyzed the MUSE spectra of M83 SNRs and a sample of \hii\ regions, we found that fitting \SiiLL\ as a doublet  and the \ha\,-\,\NiiLL\ complex as a triplet with fixed separations and a single FWHM (plus a constant background) produced more consistent results than fitting all of the lines individually, especially for the fainter spectra.  There were no cases where our visual inspections indicated any significant difference in the shapes of the \NiiLL\  compared to \ha, except those that could be associated with a narrow component of \ha\ superposed on the broader profile of the SNR.  The results of the spectral fits are presented in Tables 3and 4 for the SNRs and \hii\ regions in the sample, respectively.

% Fig. 2
\begin{figure}
\plotone{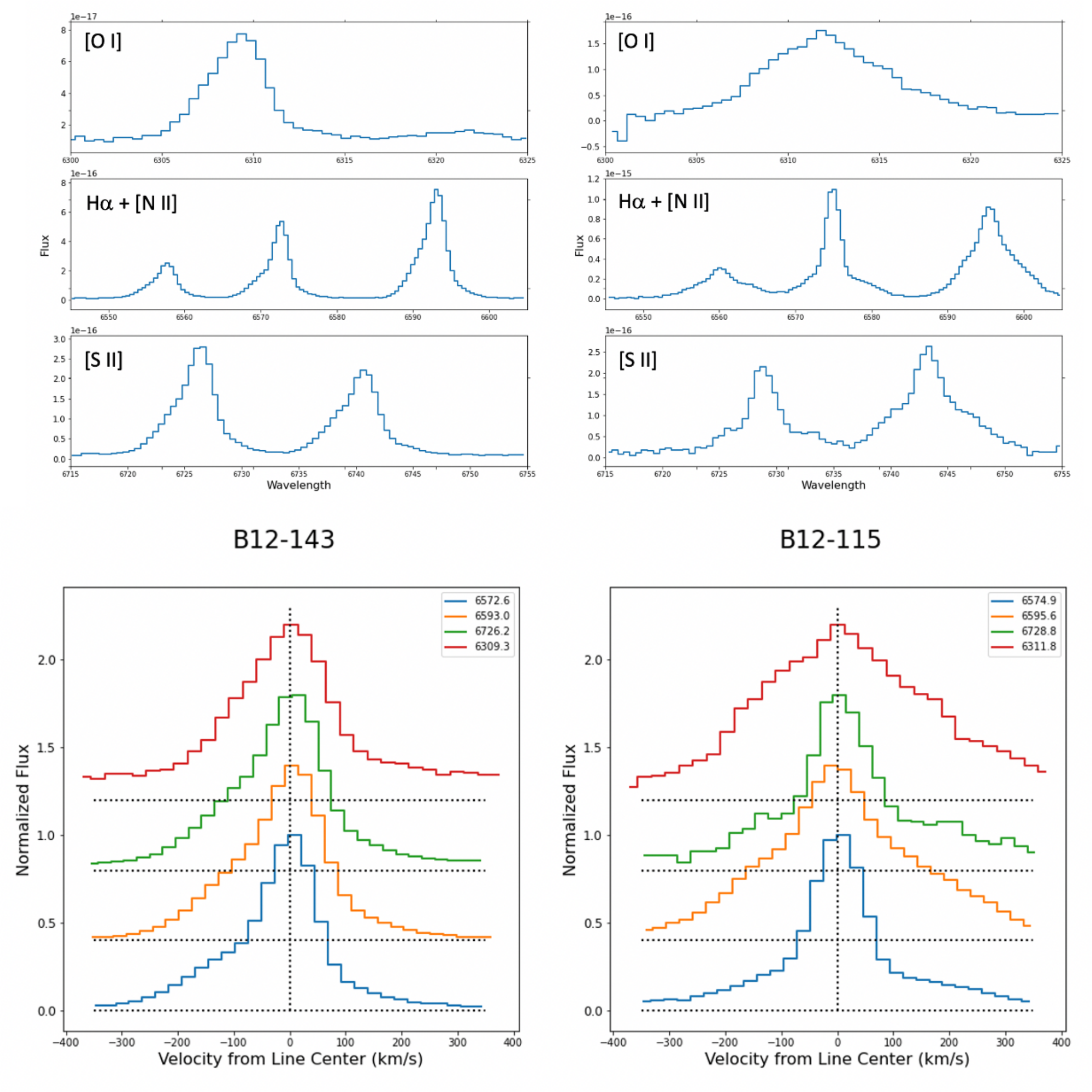}
%\plottwo{figures/B12-027_vplot.pdf}{figures/B12-075_vplot.pdf}
\caption{Similar to Fig.~\ref{fig:example_spectra} but showing the line profile changes between two SNRs.  The B12-143 spectra (left) show a clear asymmetry skewed to shorter wavelengths while B12-115 shows a broad but fairly symmetrical profile.  Note how the relative strength of the narrower line core relative to the broad component varies among the various lines plotted---especially for B12-115.   The narrow core is strongest in \ha, where narrow emission from coincident  underlying material has probably been under-subtracted.
\label{fig:example_snrs}}
\end{figure}

\subsection{Asymmetries}

% Following section on asymmetries added by WPB
The Gaussian fits provide a convenient and representative way of assessing the characteristics of the sample as a whole, and clearly show broadening for many of the SNR candidates.  However, for the first time in M83, we have sufficient spectral resolution to see that there are significant asymmetries in the line profiles of some objects. Referring back to Fig.~\ref{fig:example_snrs} (left) the spectrum of B12-143 shows that the broad component most easily identifiable with the SNR is skewed toward the blue with respect to the narrower line center.  We see other spectra where the broad component is skewed toward the red.  This is understandable as the resolved images of M83 SNRs often show shells or partial shells of clumpy emission. Since we are measuring effectively ``global" spectra of these objects, the brightest radiative shocks will dominate the emission and can shift the apparent centroid of the broad line.  In all cases, however, the broad profiles of the various emission lines in an object track each other quite well, the primary variable being the variation in strength of any narrow component emission relative to the broad component, as seen in B12-115 (Fig.~\ref{fig:example_snrs} right). Nevertheless, it is clear that the single Gaussian fits to such profiles are sufficient to capture the fact that the emission is significantly broadened, thus confirming the shock-dominated nature of each object.

As to the narrow components themselves, it is not always clear whether they are intrinsic to the SNR (e.g., asymmetric material expanding in the plane of the sky and hence near zero velocity) or due to under-subtraction of possible overlying emission.  In this regard, the \oi\ line profile may provide a useful reference (at least in cases with sufficient signal-to-noise in this weak line) since very little \oi\ is expected from overlying general emission.  In the case of B12-115, the shape and extent of the broad lines are very consistent with the \oi\ profile, likely indicating a small amount of narrow line residuals in the other lines.  Once again, these sorts of residuals have little if any significant effect on the single Gaussian line fits, which are dominated by the broad components.

To verify that any residual narrow components do not cause significant complications, we have inspected the six objects that have the most noticeable differences between the FWHMs fit to the \ha-\nii\ region and the \sii\ doublet.  These objects include B12-195 (1.8), B12-199 (1.6), B14-61 (1.3), B12-098 (1.1), B14-49 (1.1), and B12-048 (1.0), where the numbers in parentheses indicate the difference in FWHM (in \AA) between the \ha-\nii\ complex and the \sii\ doublet.  All six object spectra show evidence of a narrow component that is most conspicuous at \ha\ and at a lower level for \nii\ $\lambda$6583. Sometimes a smaller narrow component is visible for the \sii\ lines, but more typically not. The \oi\ $\lambda$6300 line never seems to show a narrow component (though the signal-to-noise is generally lower). The line ratios of the narrow component are thus consistent with what is expected for photoionized contamination of the subtracted spectrum (i.e., undersubtraction of overlying emission).

There are a relative handful of spectra where emission at higher velocities than indicated by the Gaussian fits is present at a significant level (i.e., where emission is observed in the far wings in excess of the Gaussian profile that fits the bulk of the emission).  This indicates the presence of fainter but higher velocity gas that is not being captured by the Gaussian fits.  This is to be expected as our spectral apertures no doubt sample the entire range of densities and shock velocities present; a higher velocity/lower emissivity portion of the shock would account for any such low level broad wings.  The Gaussian fits are certainly providing relevant information about the kinematics of each object, but may not always capture the full complexity and total velocity range present.

\subsection{Comparison to Previous  Spectra}
\label{sec:comparisons}

\begin{figure}
\gridline{\fig{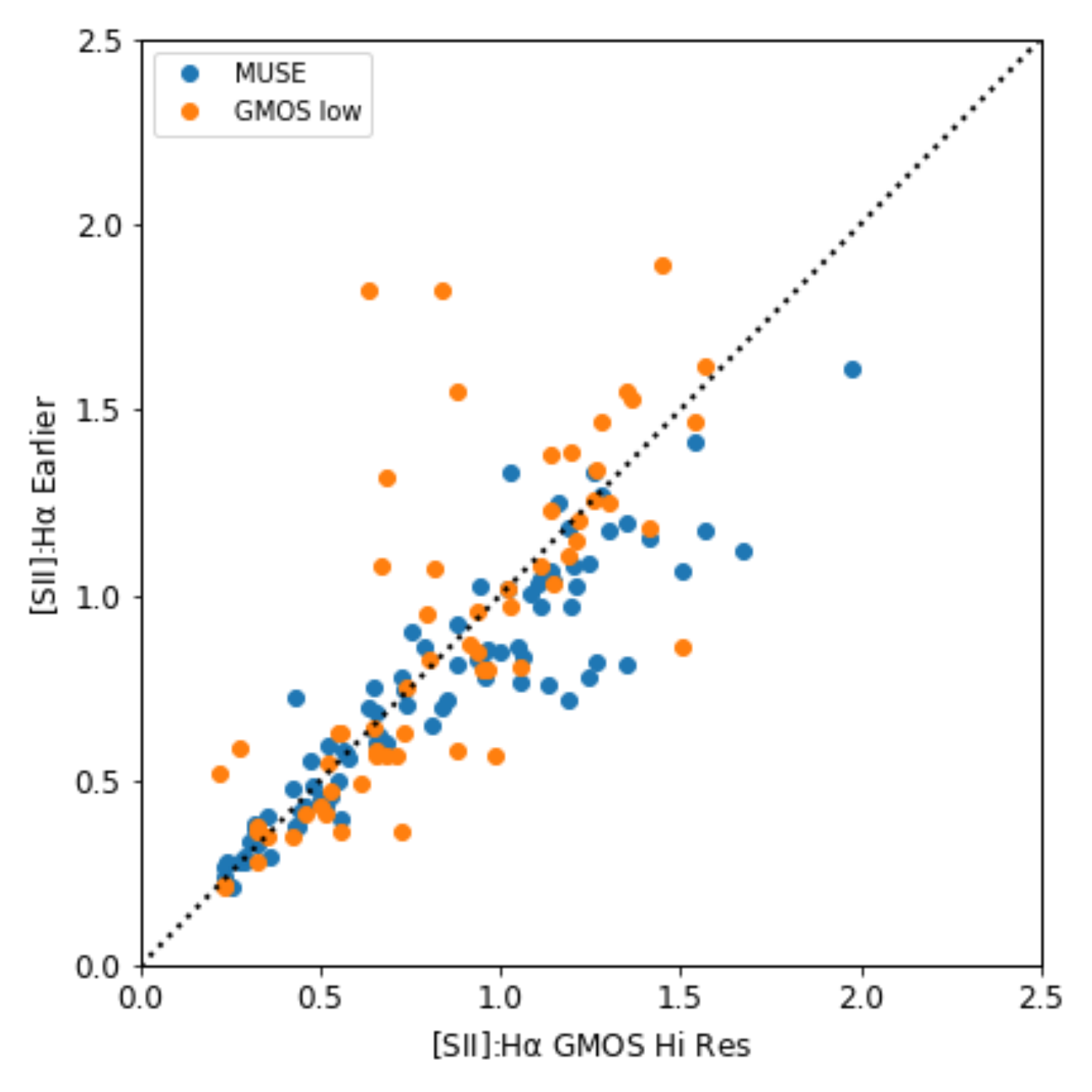}{0.33\textwidth}{(a)}
          \fig{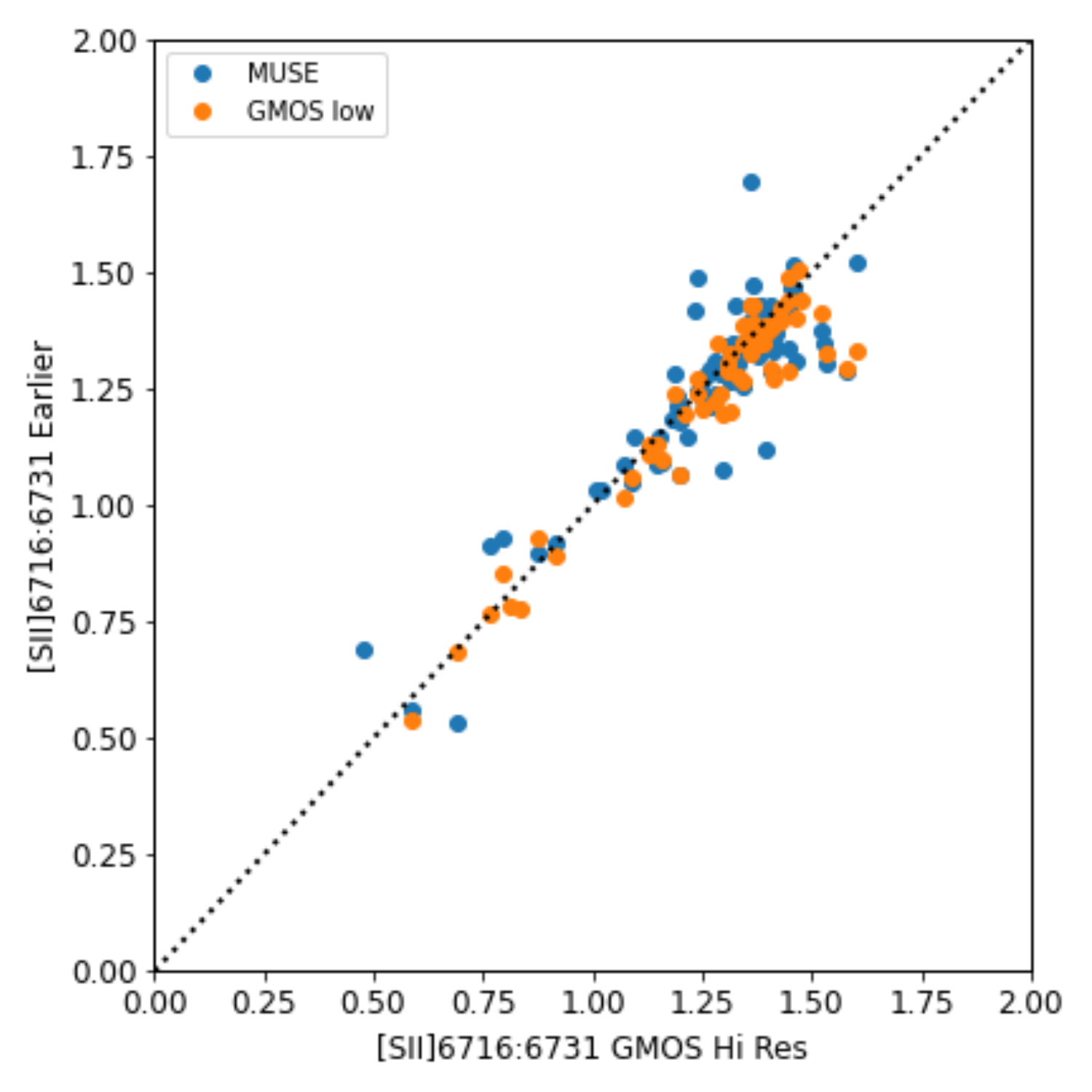}{0.33\textwidth}{(b)}
          \fig{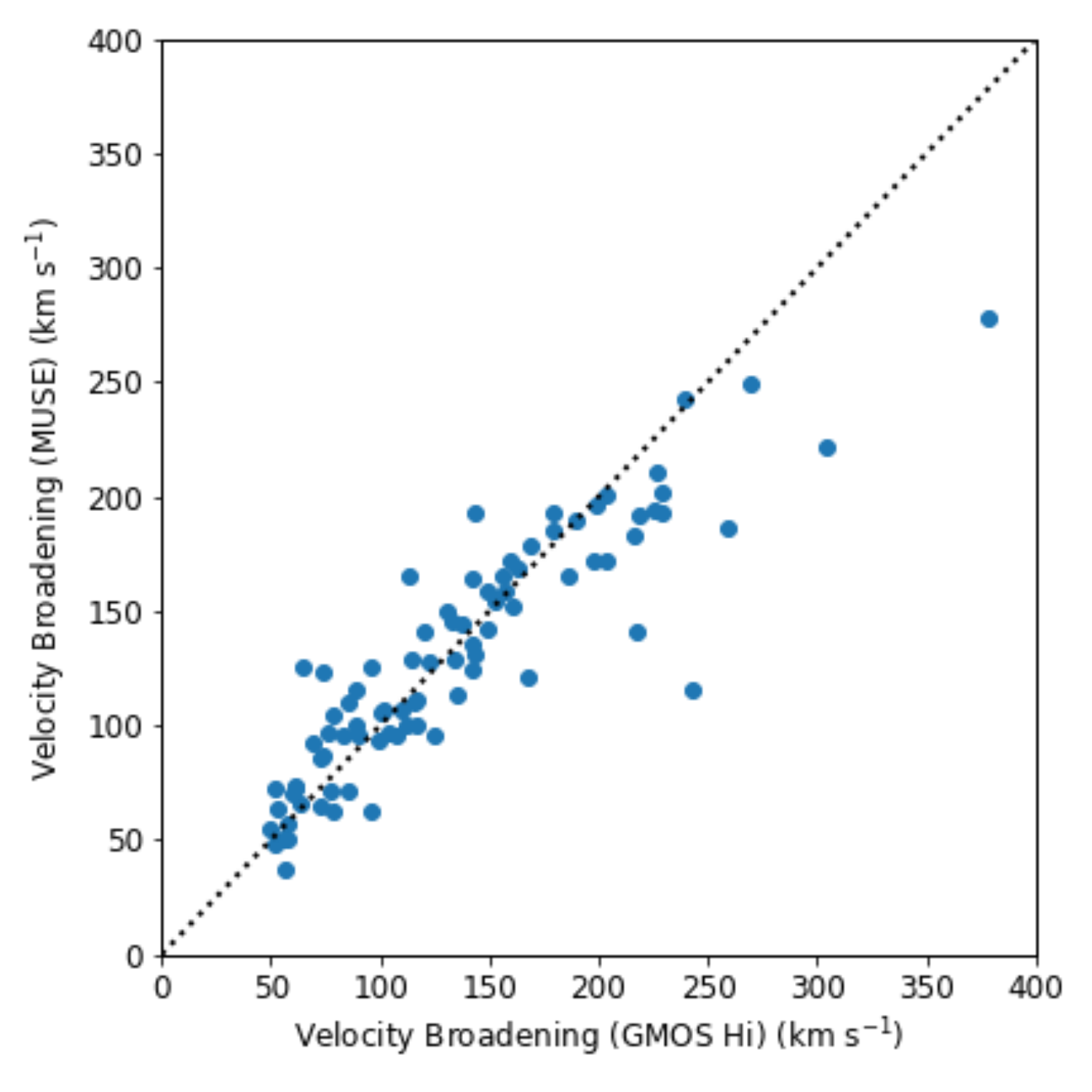}{0.33\textwidth}{(c)}}
\caption{(a) A comparison of the GMOS high resolution \sii:\ha\ ratio with earlier studies using GMOS at lower resolution \citep{winkler17} and MUSE on the VLT \citep{long22}.
(b) A similar comparison but for the \sii\,$\lambda$6716/$\lambda$6731 ratio. 
(c) A comparison of the measured velocity widths for the \ha\,-\,\nii\ complex for 
high-resolution GMOS and MUSE spectra, after allowing for differences in  intrinsic resolution of  the two instruments. \label{fig:compare} 
 } 
\end{figure}

We previously reported spectra from many of the SNRs and candidates from our studies using GMOS-S at lower resolution \citep{winkler17} and/or MUSE on the VLT \citep{long22}.  There are 58  SNR candidates in our new high-resolution set of GMOS-S spectra that had previously been observed with GMOS-S at lower resolution and 75 that had been observed with MUSE.  Our new observations include the first spectra for 24 objects in the outer galaxy that are not within the region observed with MUSE.  We compare the \sii:\ha\ ratios and the ratio of the two \sii\ lines and the velocity broadening (MUSE and GMOS high resolution observations only) for  these studies in Fig.~\ref{fig:compare}. For the velocity broadening, we have assumed that the measured widths are given by the intrinsic velocity broadening plus the instrumental resolution, added in quadrature.  Given that  different methods for background sky-subtraction were used to obtain the source spectra, there is reasonable agreement between all three sets of observations, suggesting that 
we should be able to use the full set of spectra to characterize SNRs and SNR candidates in M83.   

% Fig. 3

On the other hand,  it is also clear that the differences between the the results are larger than the statistical errors associated with fits for the values shown in Fig.\ \ref{fig:compare} and Table 3.   There are a variety of reasons for this, but the most important are likely to be (a) that the regions sampled for the sources were not identical, and (b) that different background regions and methods were used for sky subtraction.   The errors we use are the 1$\sigma$ errors derived directly from the fits; the data points do not have additional errors assigned, either due to Poisson noise of the spectra or of the background.  

Both sets of GMOS spectra were taken with slits that, in the case of the more extended sources, do not necessarily cover the complete sources, while the MUSE spectra were extracted using a circular aperture allowing for the size of the source and the seeing.  For the GMOS spectra, the backgrounds were selected from along each object's slit by inspection of the of cross-cuts of the the 2-D spectra, though the high-resolution spectra used significantly narrower slits than the low-resolution work.  For the MUSE spectra the backgrounds were selected less locally but more systematically, by locating the position within a small region around each SNR with the lowest \ha\ flux (see \citet{long22}). The background subtraction problems are  the most severe near the center of the galaxy, where the general background gas emission is especially bright.  Especially for the fainter sources, the background subtraction process using these different methods can remove different amounts of the raw flux, resulting in somewhat different relative line intensities in the subtracted spectra.  

If we take the typical systematic error between two sets of observations to be the median of the absolute value of the difference between two measurements of the same source, then the typical error in the measured line widths is 13 $\VEL$ for our new observations compared to those obtained with MUSE; if we instead take the standard deviation, it is 30$\VEL$.   Similarly, the median (STD) error for the \sii:\ha\ ratios is 0.073 (0.16) between MUSE and our new observations, and 0.084 (0.29) between our earlier GMOS observations and our current ones.  Finally, for the density-sensitive \sii\ line ratio, the median (STD) error is 0.031 (0.1) for the MUSE vs. our current observations, and 0.035 (0.07) for the earlier GMOS vs. the current GMOS observations.  The fact that these cross-comparisons show considerably larger errors than the statistical errors explains why we do not make extensive use of the statistical errors in the discussion below.

\subsection{Kinematics as an Independent Indicator of SNR Shocks}

\begin{figure}
\gridline{\fig{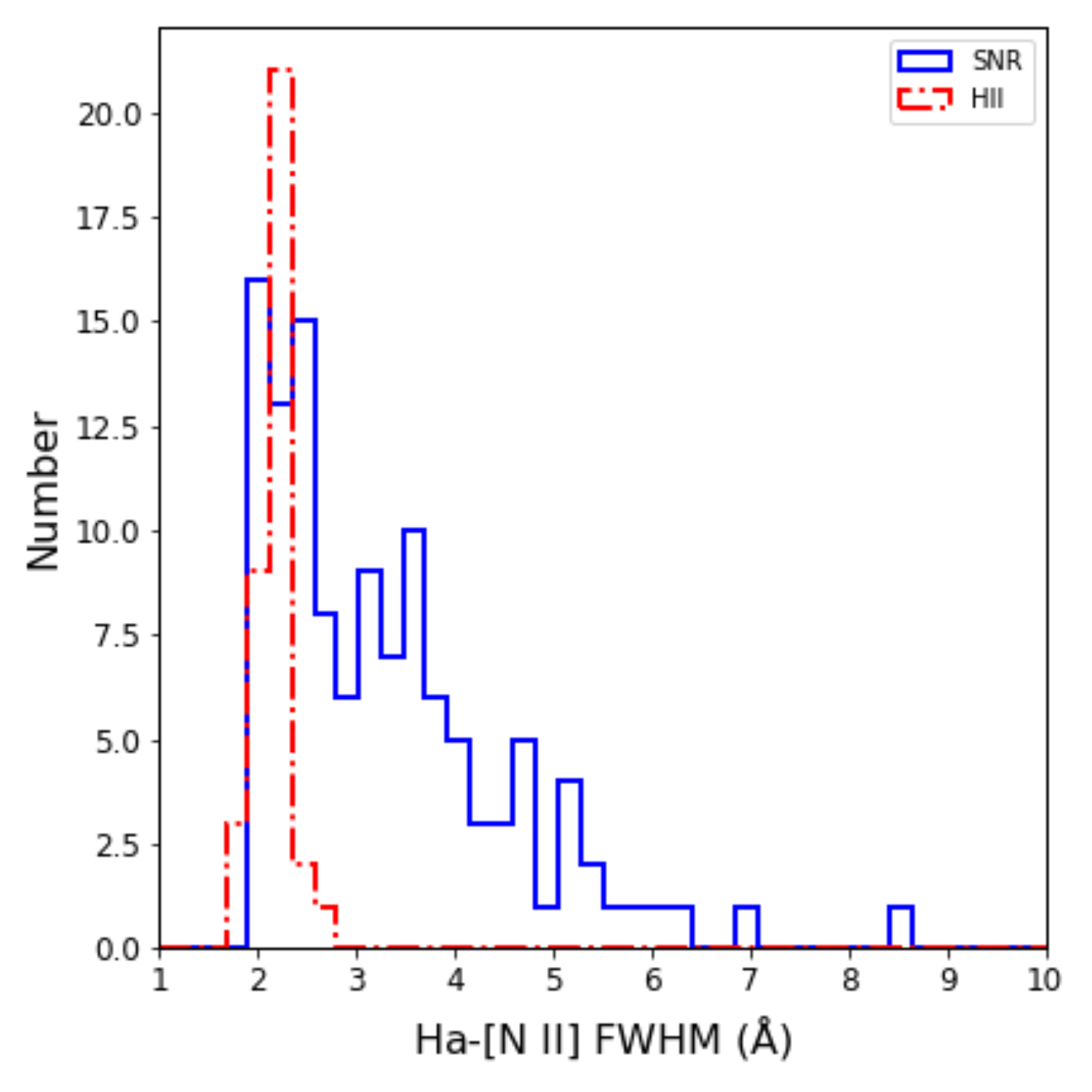}{0.33\textwidth}{(a)}
          \fig{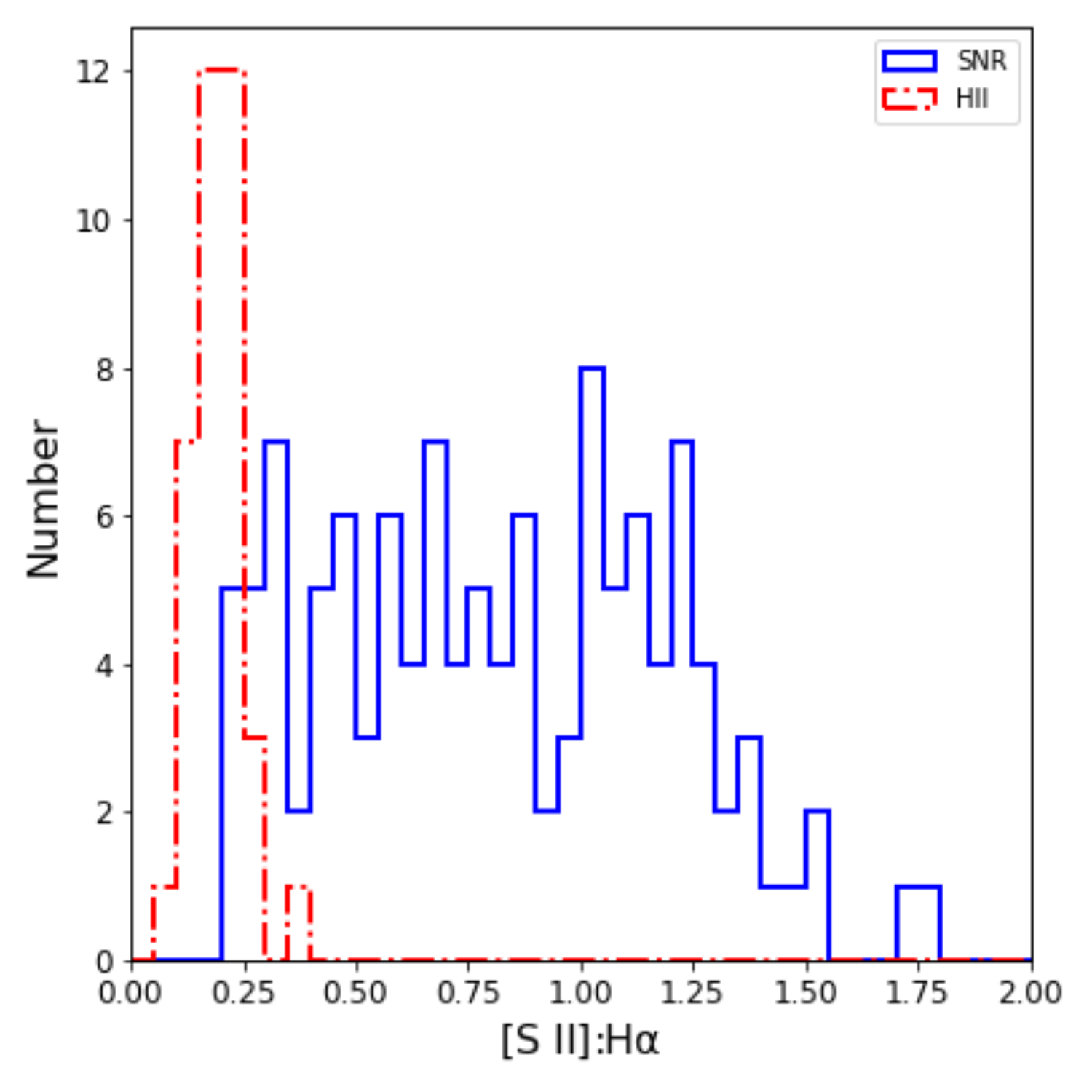}{0.33\textwidth}{(b)}
          \fig{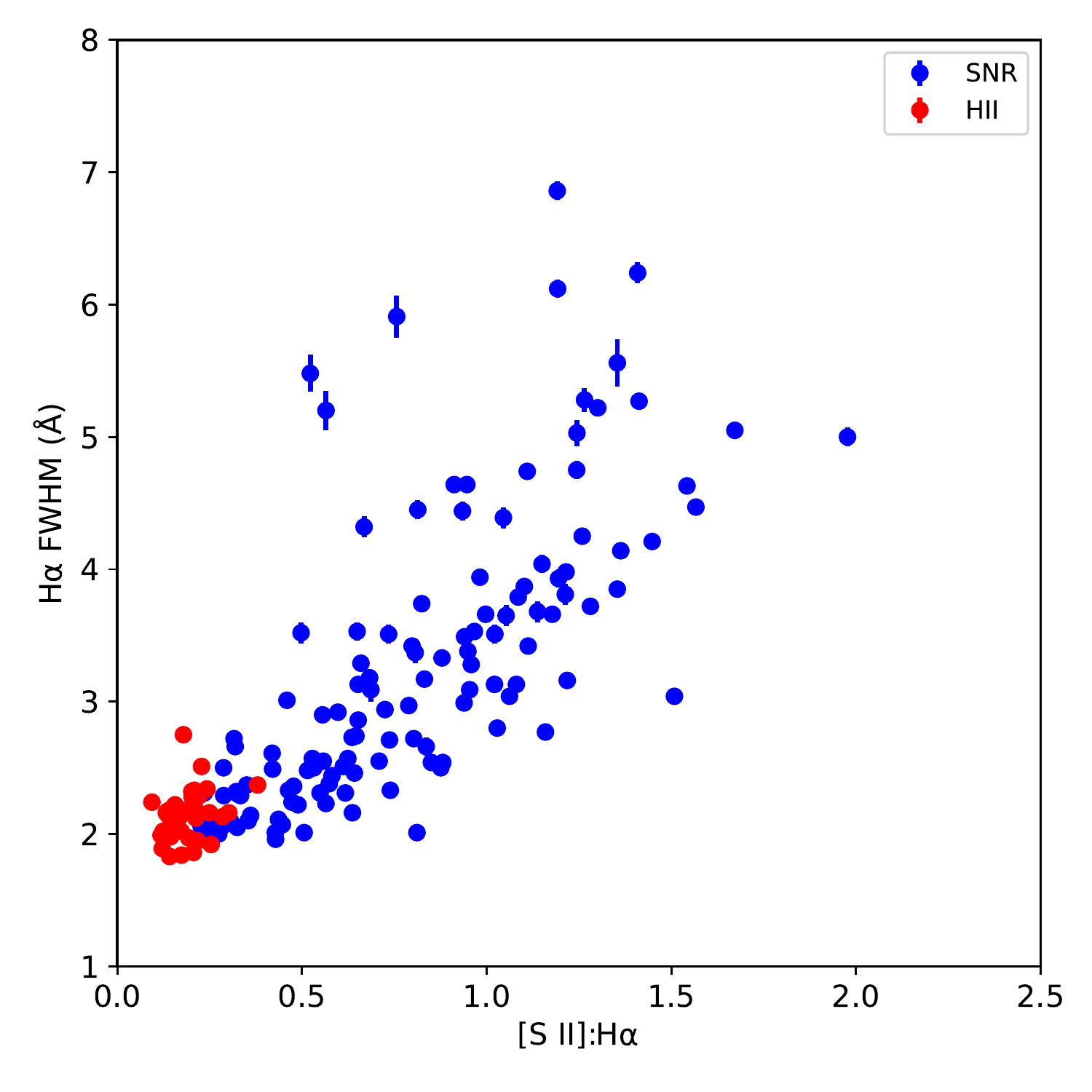}{0.33\textwidth}{(c)}}
\caption{(a) A comparison of the distribution of the measured FWHM of the \ha-[N~II] lines for  the SNRs and SNR candidates to the \hii\ regions in the current study.
(b) A similar comparison but for the [S~II]:\ha\ ratio.
(c) The measured FWHM of the lines for \hii\ and SNR candidates as a function of the \sii:\ha\ line ratio.
\label{fig:hist}
 } 
\end{figure}

One of the primary reasons we proposed to carry out higher resolution GMOS observations of the SNR population of M83 was to use velocity widths as an independent diagnostic to determine whether candidate SNRs were indeed SNRs.  
Considering only the new higher resolution GMOS spectra first, in
Fig.\ \ref{fig:hist}, we show the measured properties of the SNRs and SNR candidates and the \hii\ regions are generally quite different in terms of the FWHM of the lines, and of the line ratio for \sii:\ha. In addition, we see a fairly strong correlation between the \sii:\ha\ ratio and the FWHM.  If we take the intrinsic resolution of the spectra as 85 $\VEL$, then an intrinsic velocity spread of 100$\VEL$ added in quadrature corresponds to a FWHM of 2.87 \AA\ at \ha.  Of the 36 \hii\ regions with new GMOS spectra, none has FWHM greater than this, while 69 (58\%) of the SNR and SNR candidates do.  If we include the objects with MUSE spectra as well, we find 159/272 objects show FWHM $>$100$\VEL$, also 58\%.

These results make it quite clear that higher resolution spectra of SNRs are a useful tool for independently distinguishing SNRs from \hii\ regions, as previously pointed out by \cite{points19} for SNRs in the LMC\@. 
However, the results also make clear that there is a region of overlap at the lowest FWHM end where SNR candidates and \hii\ region spectra are essentially still unresolved at our resolution. The SNR candidates in this region might have low enough bulk motions so as not to produce observed broadening even at the increased spectral resolution employed here. Given that bulk velocities in \hii\ regions are typically 10-20  $\VEL$, the primary lesson may be that if one is to use velocity broadening for identifying SNRs, one needs higher spectral resolution  to separate ``all'' of the SNRs from \hii\ regions.  The evident correlation between the  \sii:\ha\ ratio and the FWHM is most likely a physical effect associated with shock physics, a topic to which we will return in  Sec.\ \ref{sec:discussion} below.

As a result, we conclude that having  velocity broadening $> 100\, \kms$ is almost certainly a {\em sufficient} condition to confirm an object as an SNR, but it is not a {\em necessary} one.  Perhaps this should not surprise us; \hii\ regions typically have velocity dispersion $\lesssim 20\, \kms$.  Our spectral resolution of $85\, \kms$ appears not sufficient to measure the velocity broadening of the most evolved SNRs, whose shock velocities may be lower than younger SNRs.  Unfortunately, no instruments with higher spectral resolution and multi-object capability are available at the Gemini Observatory.

 % Fig. 4

\subsection{Young SNRs}

Young remnants of core-collapse supernovae are characterized by very broad emission lines of O, and possibly other elements, Two such remnants are known in M83: SN1957D and B12-174a \citep[the remnant of an event that was not observed, but that almost certainly took place within the past century;][] {blair15}.   Spectra of both were included in our high-resolution GMOS program.  The spectrum of B12-174a (taken in 2021) is very similar to the original one from 2011 that appears in \citet{blair15}.  The higher resolution of the more recent spectrum does not reveal any new features, since the lines that characterize this spectrum have velocity width $\gtrsim 5000 \kms$.

For SN1957D, the 2021 spectrum does not extend far enough into the blue to reach the \oiii\ lines, which  in previous spectra \citep{long89, turatto89, long91, long12} were by far the strongest broad emission lines.  Furthermore,  the new spectrum has lower sensitivity, especially to broad lines, than our low-resolution GMOS spectrum from 2011 \citep{long12}, and it  has significantly worse signal-to-noise---especially so because this was among the mask 5 objects, for which the observations were inadvertently taken with no binning (vs 2$\times $2 binning for the other four masks).  The 2021 spectrum does suggest broad \oi\ $\lambda$\,6300, at a flux and width comparable to that seen in 2011\@. But the new high-resolution observations simply lack sufficient sensitivity and signal-to-noise to reach any definitive conclusions about the brightness of the faint broad lines or the evolution of this object over the past decade.

\section{Discussion \label{sec:discussion} }

\subsection{Overall characterization of the sample}

As noted earlier, we have carried out a series of studies of M83 intended in large measure to characterize the SNR population there. Combining the results from \cite{winkler17} and \cite{long22} with the current study, we have obtained optical spectra of 310 of 366 SNR candidates in M83, 272 of these with sufficient resolution to determine if the velocity broadening of the lines exceeds about 100 $\VEL$.  M83 is the only galaxy for which such a large number of high-resolution spectroscopic observations has been carried out.\footnote{In addition to the high-resolution GMOS observations, we previously observed five bright M83 SNRs plus three \hii\ regions with the Goodman Spectrograph on the SOAR telescope, in longslit mode, as  reported in \citet{points19}.  These observations used a 0\farcs 6 slit, and have similar velocity resolution to that of the new GMOS spectra,  All but one of these five SNRs are also among those with new GMOS spectra, and also have spectra from MUSE; hence, we have not repeated the Goodman results here.}

The  results from all of these investigations for each SNR candidate are summarized in Table 5, and the utility of the various criteria we use are summarized in Table 6. Of the 310 SNR and SNR candidates for which we have spectra, 238 have \sii:\ha\ ratios of greater than 0.4 in at least one spectrum, and even more, 249, have significant \oi\ emission \edit1{(which we take as \oi:\ha\, $>0.03$, 
the value below which the signal-to-noise in our typical spectra becomes  unacceptably low.)  \citet{kopsacheili20} have argued that a lower value of \oi:\ha\, = 0.01 cleanly separates SNRs from \hii\ regions, but our spectra cannot reliably measure such low ratios}.  We will discus this topic in further detail below. In Table 6 we include statistics for the entire galaxy, and for the extra-nuclear region (galactocentric radius $>$ 0.5\,kpc), since the source confusion and overall diffuse background in the nuclear region render background subtraction uncertain.  

If we consider only the objects with spectra that have sufficient resolution to measure line widths as small as $100\, \kms$, i.e., the MUSE and GMOS high-resolution spectra, we can compare the efficacy of all the  criteria for confirming objects as SNRs, including velocity broadening.  Again we find that almost equal numbers have \sii:\ha\ ratio greater than 0.4 and \oi:\ha\ ratio greater than 0.03; most objects which obey one of these criteria obey both.  Regarding velocity broadening, a smaller number have broadening in excess of 100 $\VEL$ (FWHM), which is the best that our data permit for separating SNRs from \hii\ regions.  Virtually all of the broadened objects also obey the \sii:\ha\ ratio or \oi:\ha\ ratio criteria, or both, as was true when we considered only the new GMOS data.  

We now know that many SNRs are detectable using the near-IR line of \feii\ $\lambda\,$1.644$\mu$m, and indeed many SNRs in NGC\,6946 have been detected using this diagnostic \citep{long20}.  Although we first published detections of M83 SNRs as sources of \feii\ emission in \cite{blair14}, using WFC3/IR with the F164N filter, that paper  reported only the subset of objects that met the other criteria for that paper, namely. small angular-size objects and SNR candidates found uniquely in HST WFC3-IR imaging data.  We have never provided an accounting of which M83 SNR candidates were first identified via their \feii\ emission as well as which optical SNR candidates displayed $detectable$ \feii\ emission at the exposure level available. 

We make up for this past shortcoming in Table 5, where the \feii\ column indicates the following: an entry of `YES' means the candidate was {\em first} identified from its \feii\ emission.  Many of these objects are located in dusty regions (including in the nuclear region) where the optical source emission is weak or absent but the \feii\ emission is visible.  An entry of `yes' indicates other SNR candidates that appear to have an \feii\ counterpart to the WFC3-VIS imagery, based on visual inspection at the location of each optical candidate. An entry of `no' indicates the object was within the footprint of the region observed with the F164N filter but for which no counterpart could be identified.  Finally, a dash indicates that the object was outside of the region observed with F164N.  The table contains 29 objects discovered using \feii, and an additional 118 objects detected in \feii, for a total of 147 \feii\ detections. We also find that there were a total of 287 objects within the region observed with F164N.  Hence, at the current exposure level in M83 the \feii\ detection rate is just over 51\%.

\edit1{ The \feii\ line is a good discriminant between SNRs and \hii\ regions partly because shocks produce low-ionization emission, and partly because shocks liberate Fe from dust. According to \citet[][Fig 7]{slavin15}, the fraction of silicate dust mass returned to the gas phase rises steeply from 0 at $50 \kms$ to 40\% at $100 \kms$. This might also create a tendency for SNRs with higher velocity widths to show up most strongly in \feii.}

We might conclude that the ``gold standard" set of objects that are certainly SNRs includes those with clear velocity broadening, \oi, \feii, {\em and} a \sii:\ha\ ratio greater than 0.4.   But by  demanding that all four criteria be met, we are certainly excluding many objects that are actual SNRs; if the goal is to get a {\em complete} sample of SNRs, choosing  only such a ``gold" sample of nebulae is far too conservative.

Instead, for the purposes of the discussion that follows, we assume that emission nebulae with a spectroscopically measured \sii:\ha\ ratio greater than 0.4 are SNRs.  As discussed in Section \ref{sec:failed},  some of the objects that have \sii:\ha\  ratios less than 0.4 may also be SNRs, but it is quite likely that this group also contains many objects that are not SNRs.

\subsection{Comparison to shock models}

% Figure 5
\begin{figure}
\plottwo{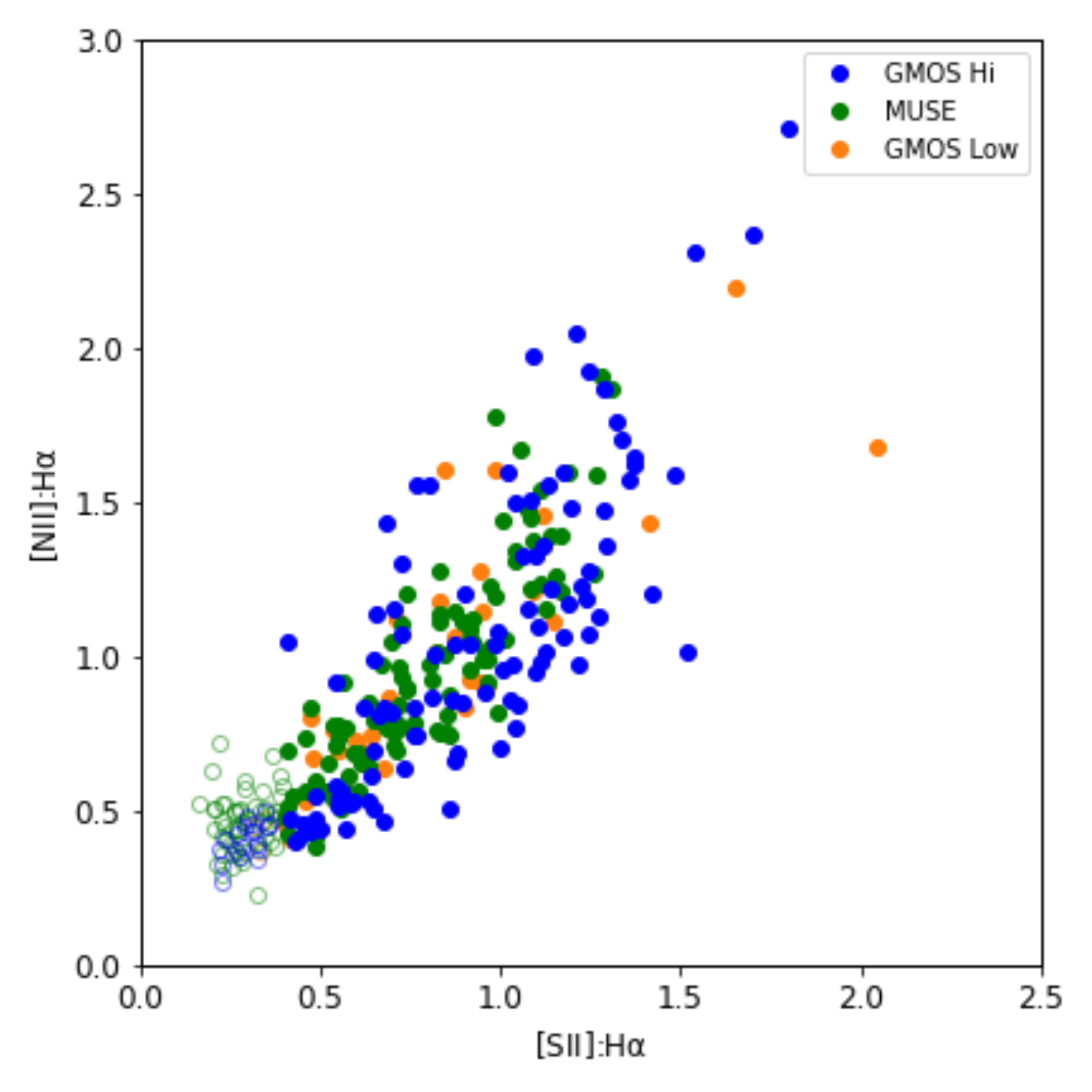}{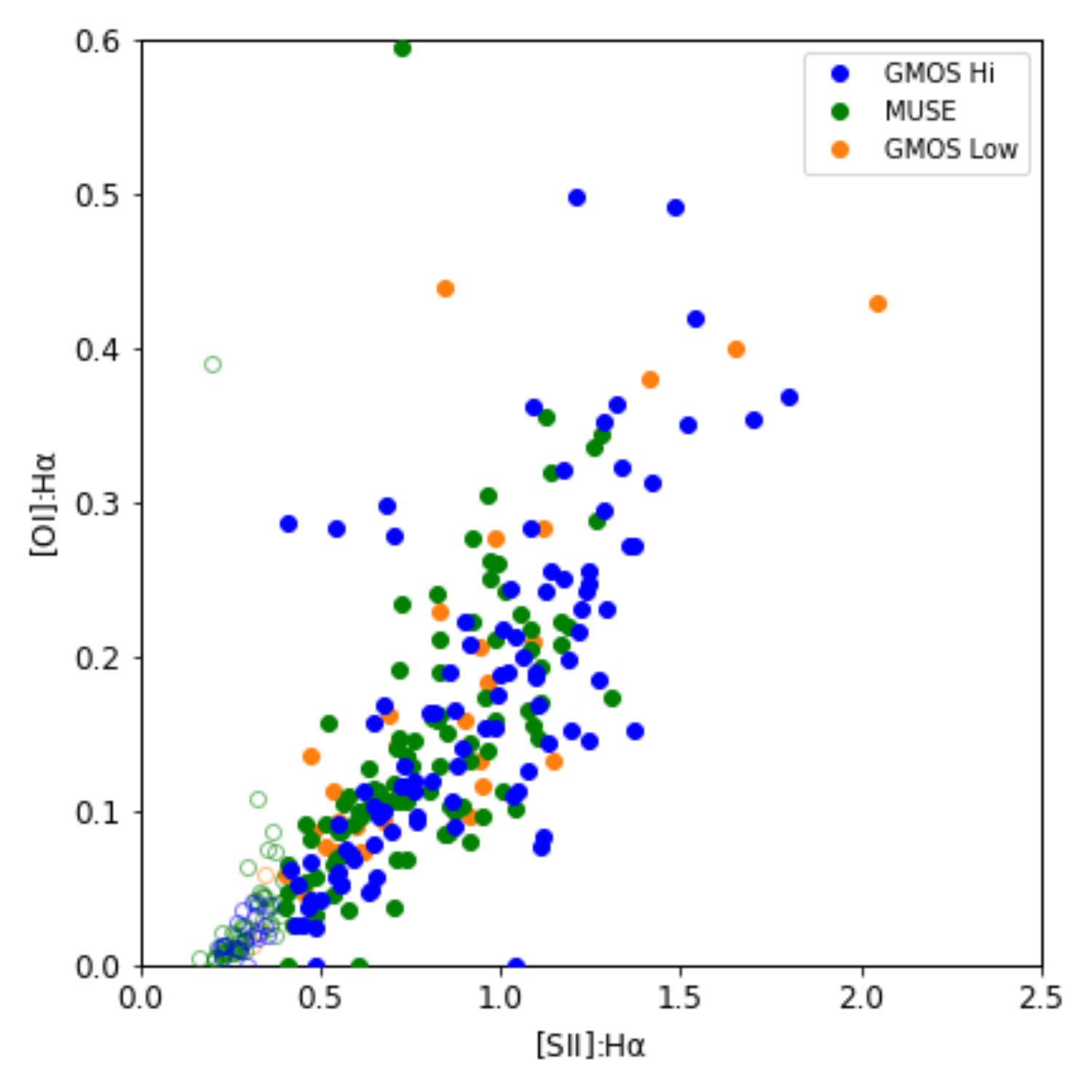}
\caption{Correlations between the \nii:\ha\ and \oi:\ha\ ratios against the \sii:\ha\ ratio. Objects with \sii:\ha\ ratios $\ge$ or $<$ 0.4 are indicated by filled and open points, respectively.} 
\label{fig:correlations}
\end{figure}

Since the early work by \cite{cox72}, the optical spectra of Galactic (and Magellanic Cloud) SNRs have usually been interpreted in the context of plane-parallel radiative shocks with shock velocities of 100 to 400 $\VEL$ expanding into a uniform medium.  In M83 and other nearby galaxies, one would like to be able to use the the spectra that have been obtained and shock models to determine the nature of each SNR and the environment into which it is expanding.  However, it is unclear that this is possible.  Firstly, the spectra of SNRs in M83 (and other similar galaxies are `globally averaged' spectra of entire objects while shock models would typically be most applicable to a specific shocked filament in a SNR that might be better approximated by a particular set of model conditions.

% Figure 6
\begin{figure}
\plottwo{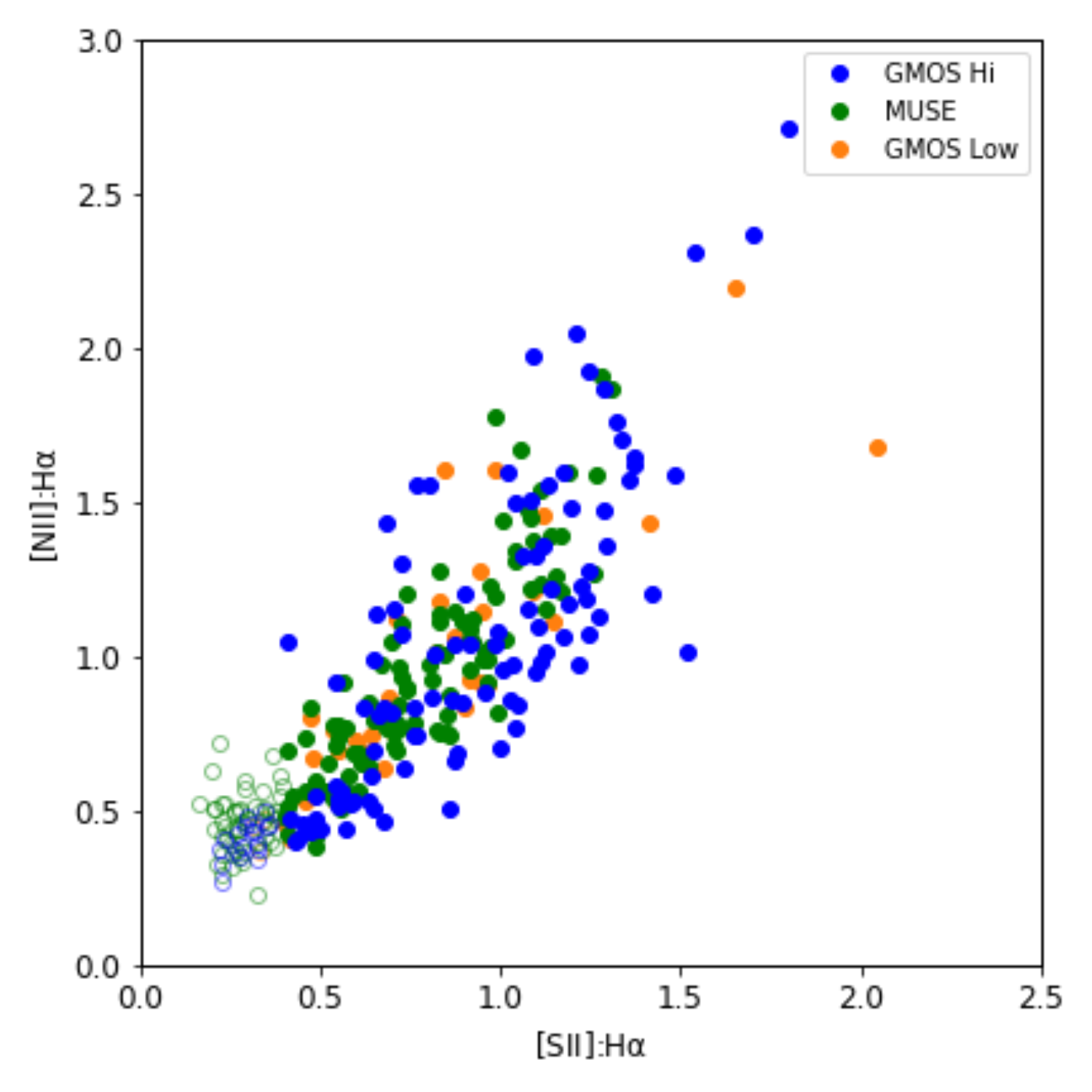}{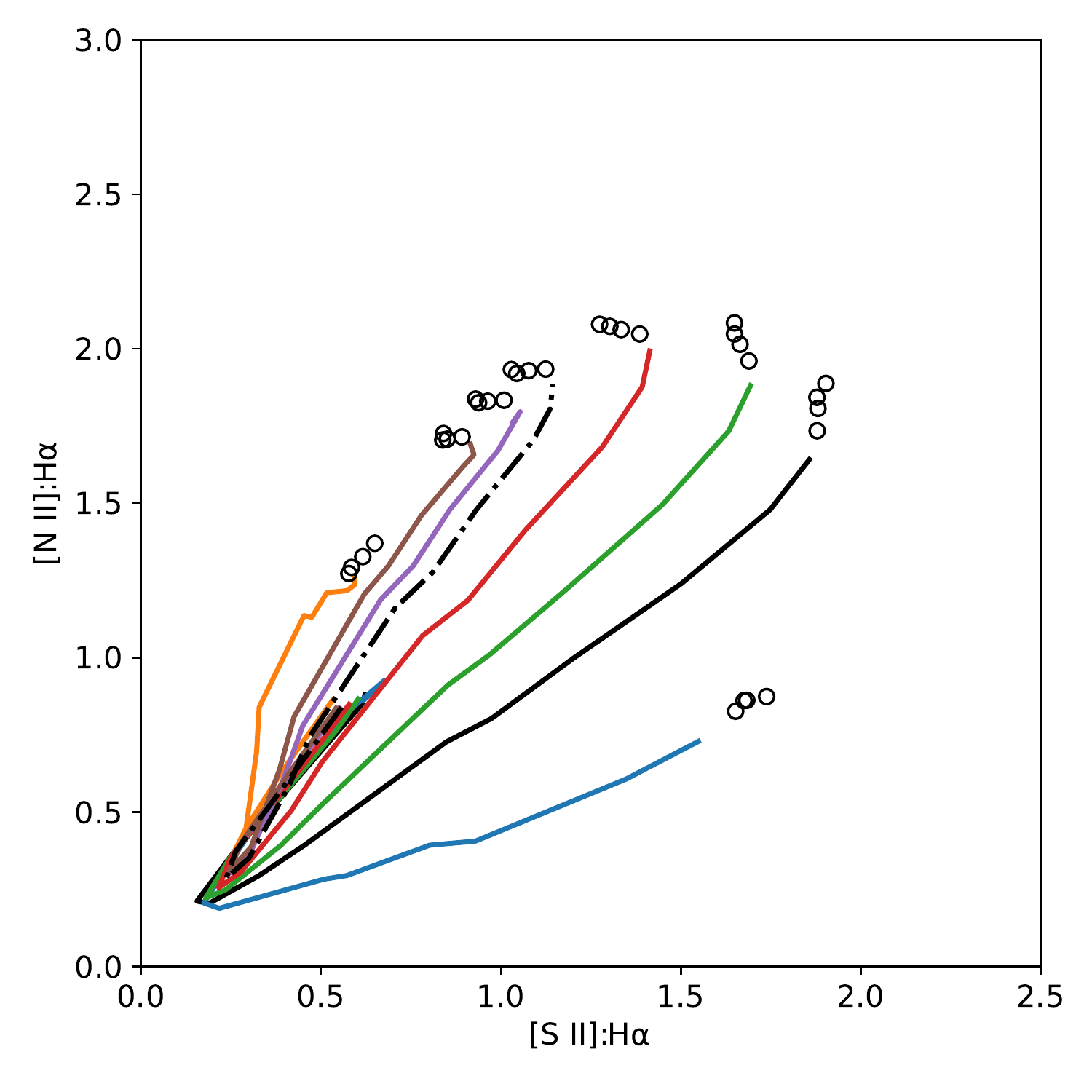}
\plottwo{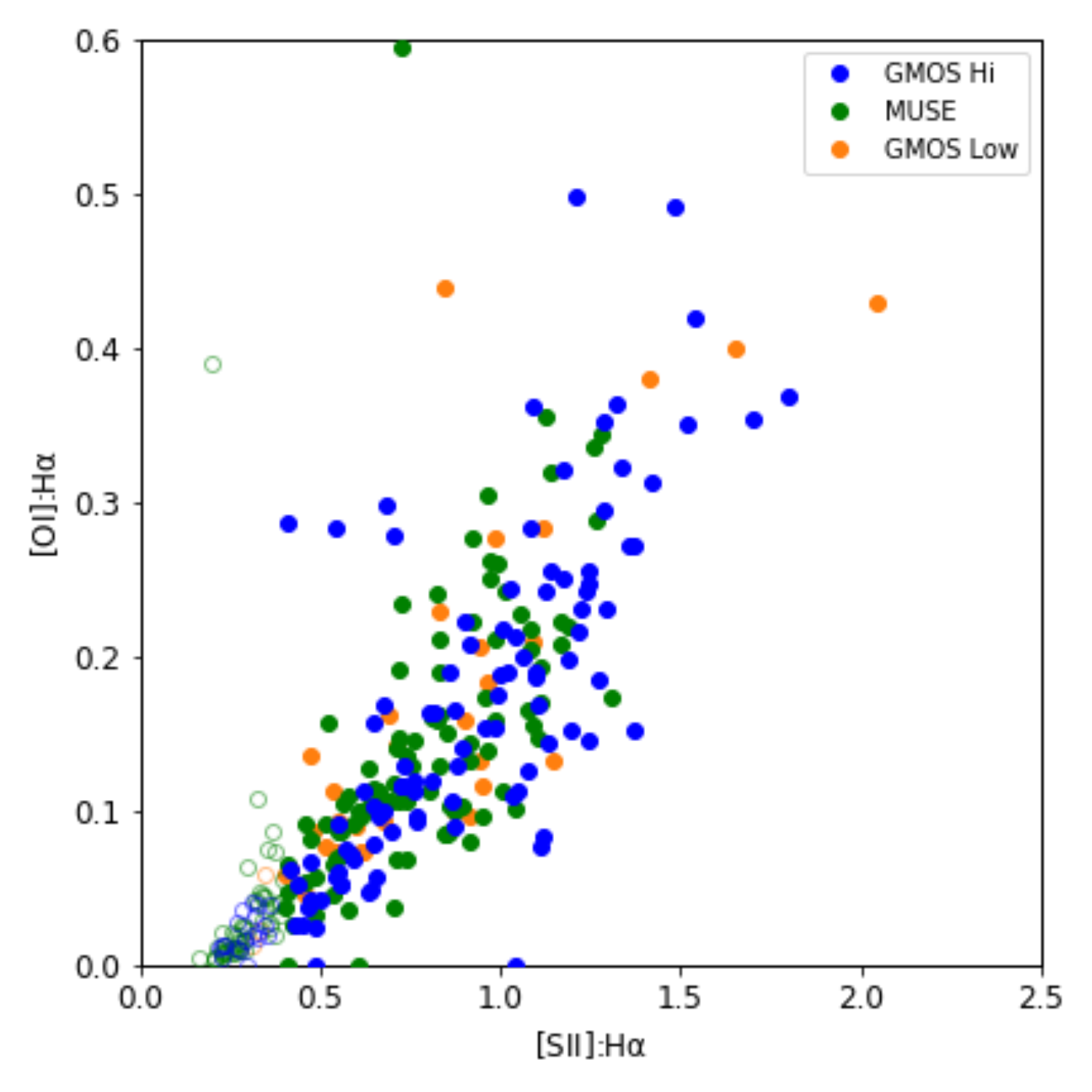}{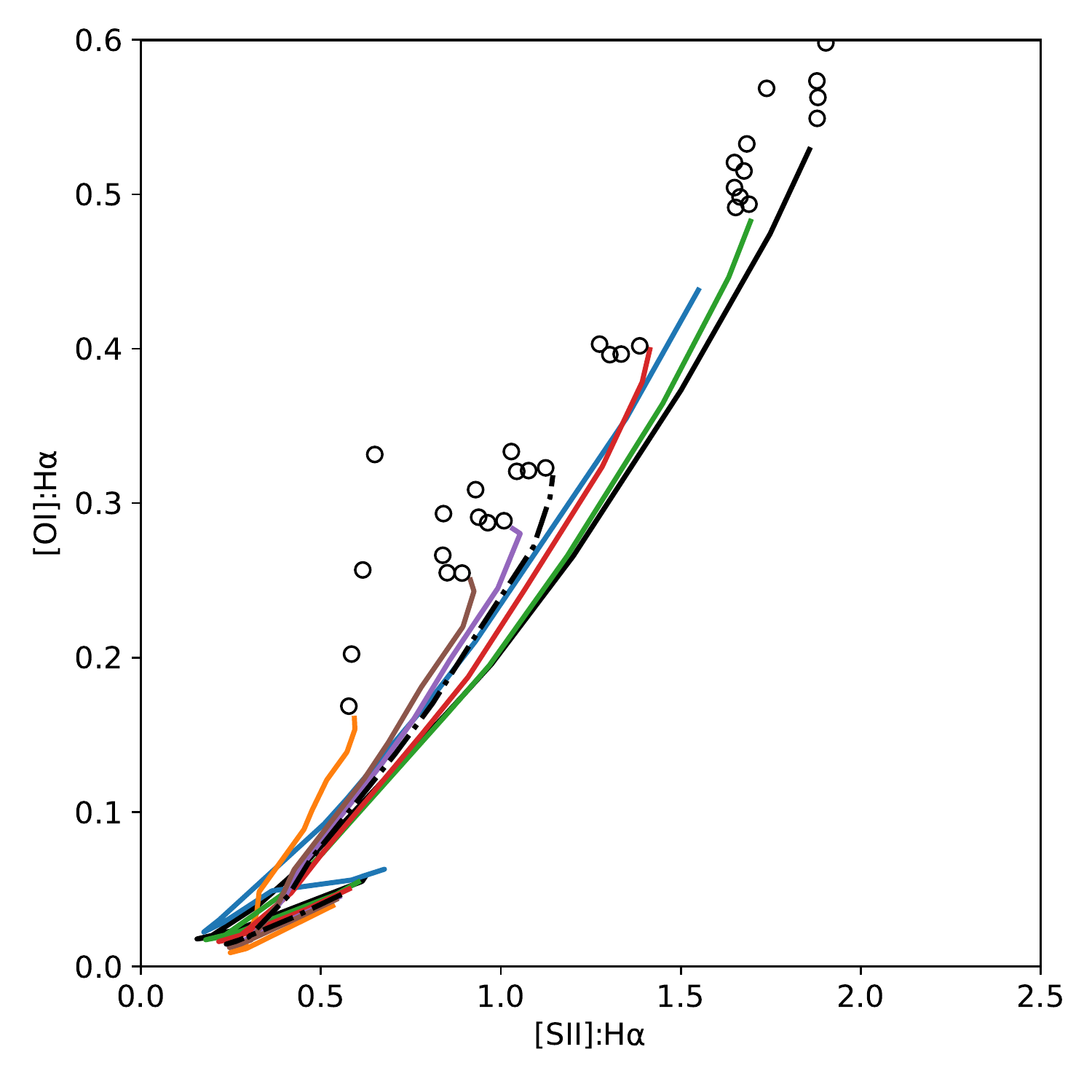}
\plottwo{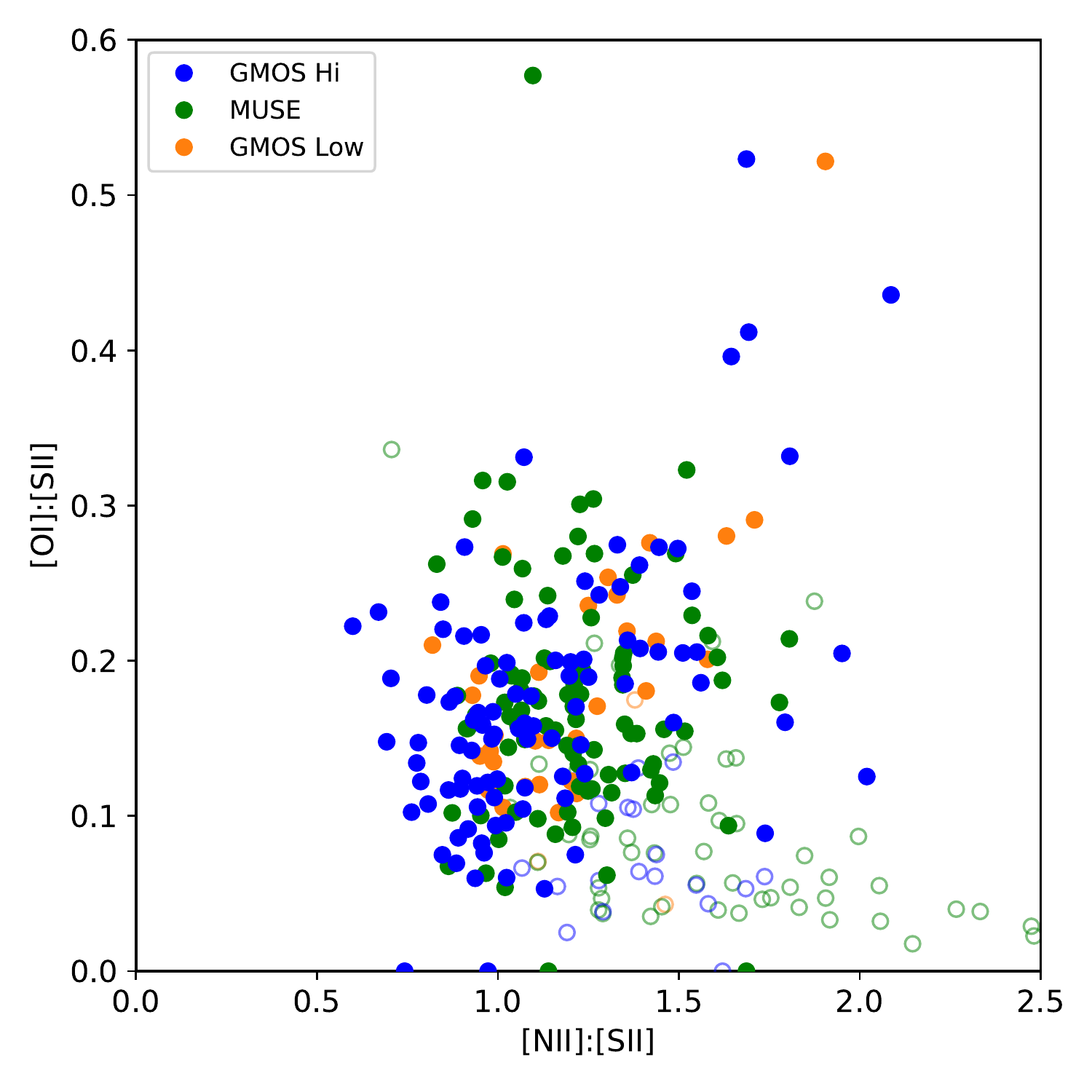}{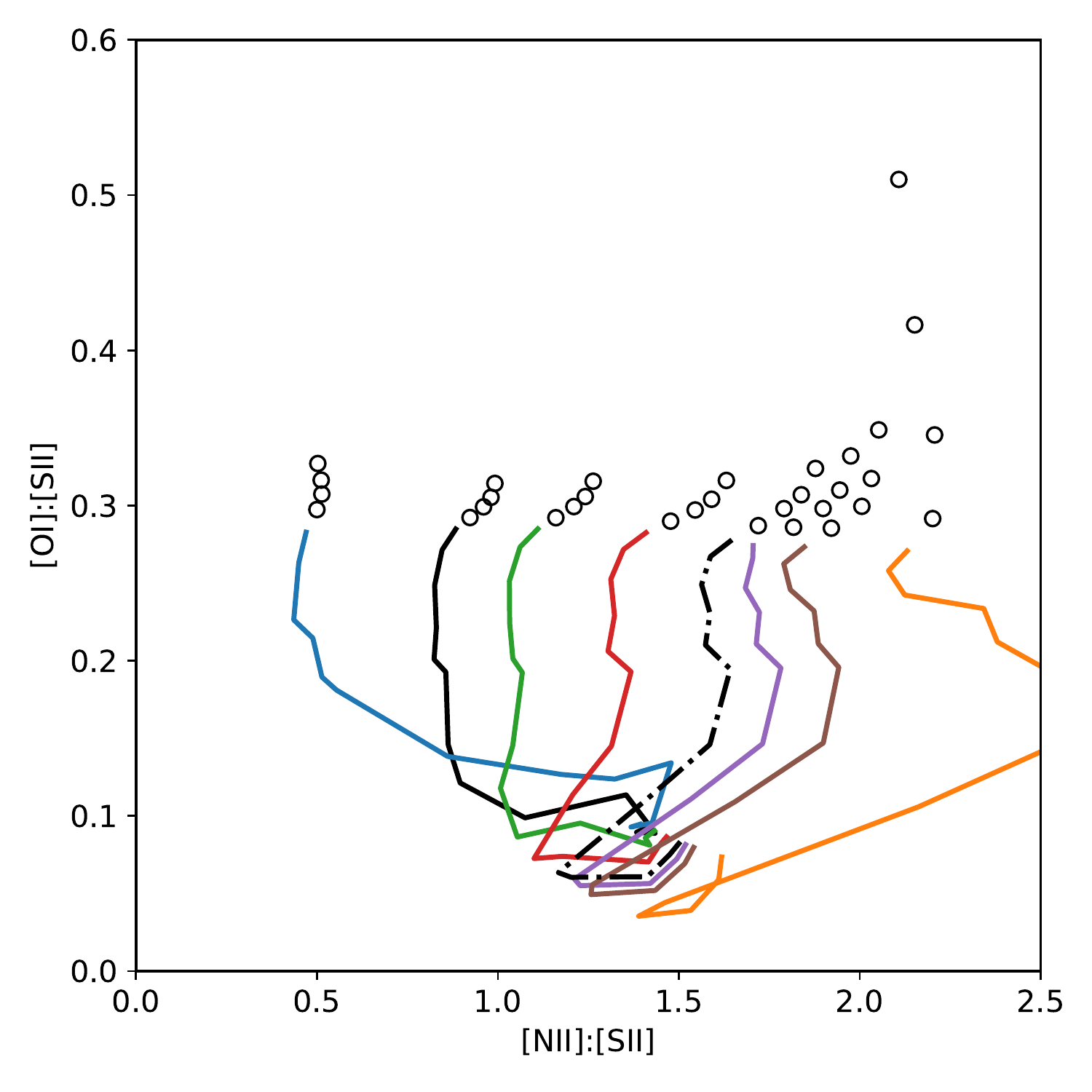}
\caption{Comparisons of the various observed line ratios to line ratios calculated by  \cite{allen08} assuming abundances that are twice solar and a pre-shock density of 1 $\rm cm^{-3}$.   The various lines are for different values of the magnetic field,\edit1{ ranging from no magnetic field (solid black line) to 10 $\mu$G (black 'dash-dot' line).} The solid lines are for velocities less than 400 $\VEL$, while the open circles represent shock models from 400 to 500 $\VEL$.  Higher values of the shock velocity are not plotted, because these have implausibly long recombination times for SNR applications.  \edit1{ The models do not form a simple progression of line ratios with magnetic field strength.  See \cite{allen08} for details.} }
\label{fig:model_compare}
\end{figure}
%\includegraphics[]{figures/fig_s2_den_vel.pdf}

% Fig. 7
\begin{figure}
\gridline{\fig{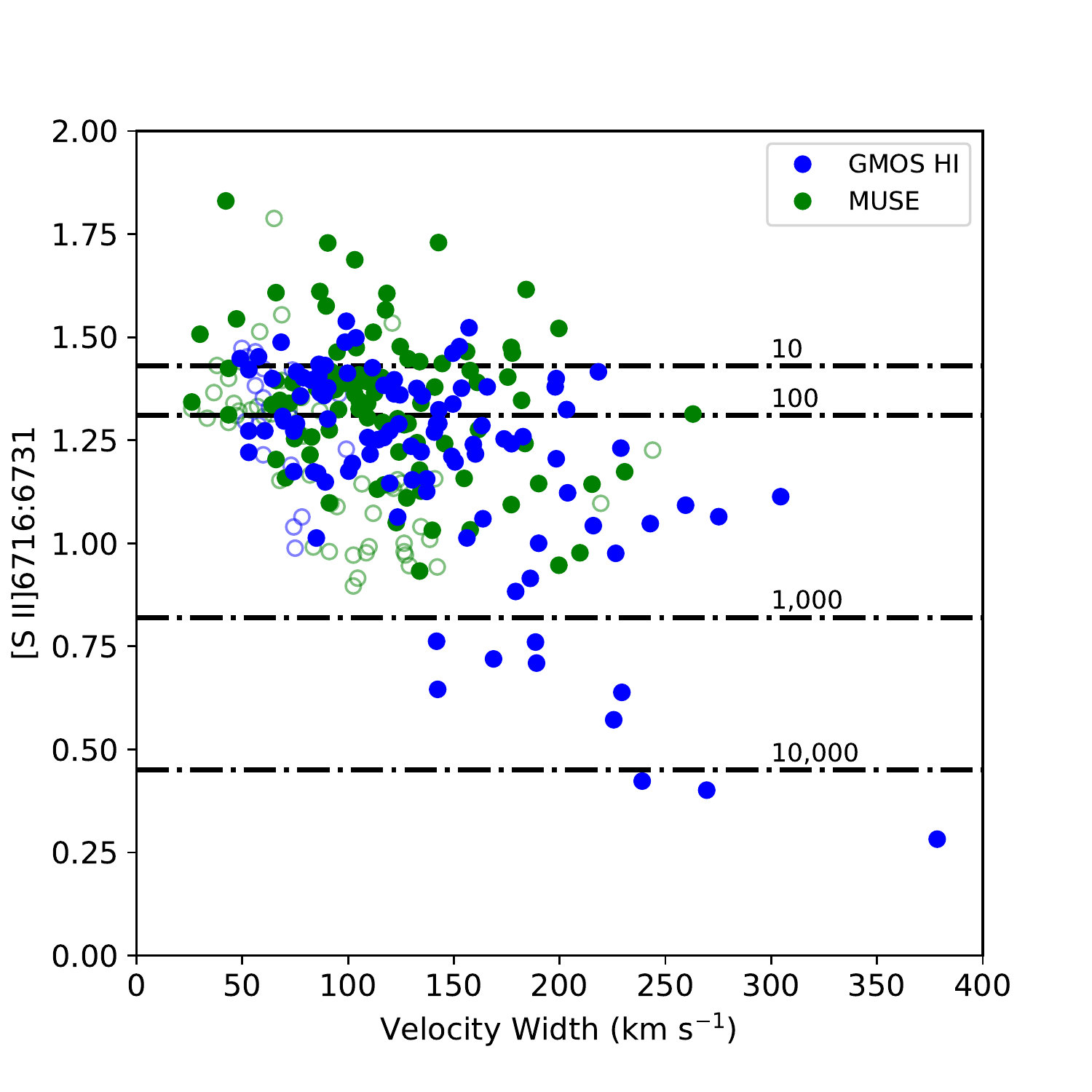}{0.33\textwidth}{(a)}
          \fig{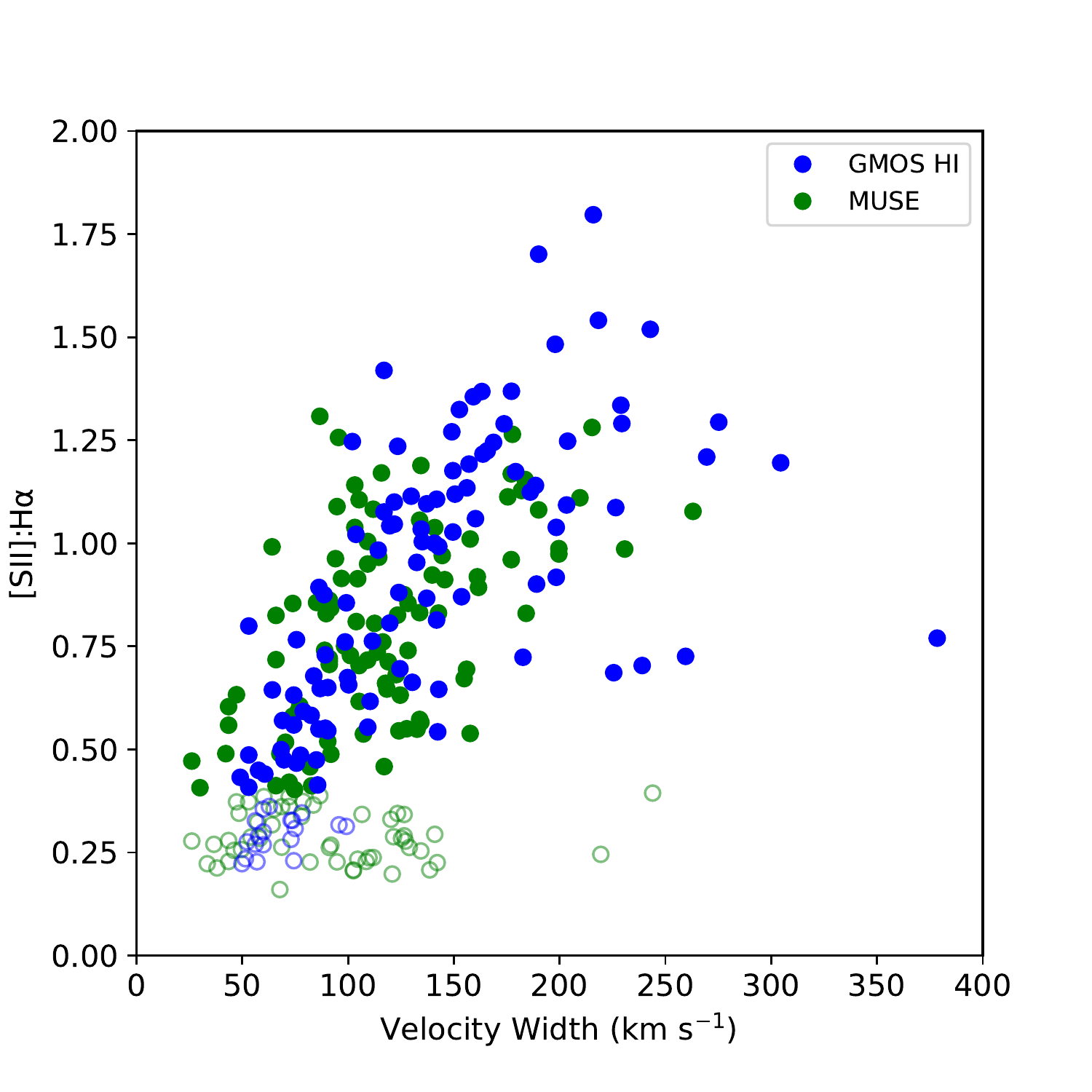}{0.33\textwidth}{(b)}
          \fig{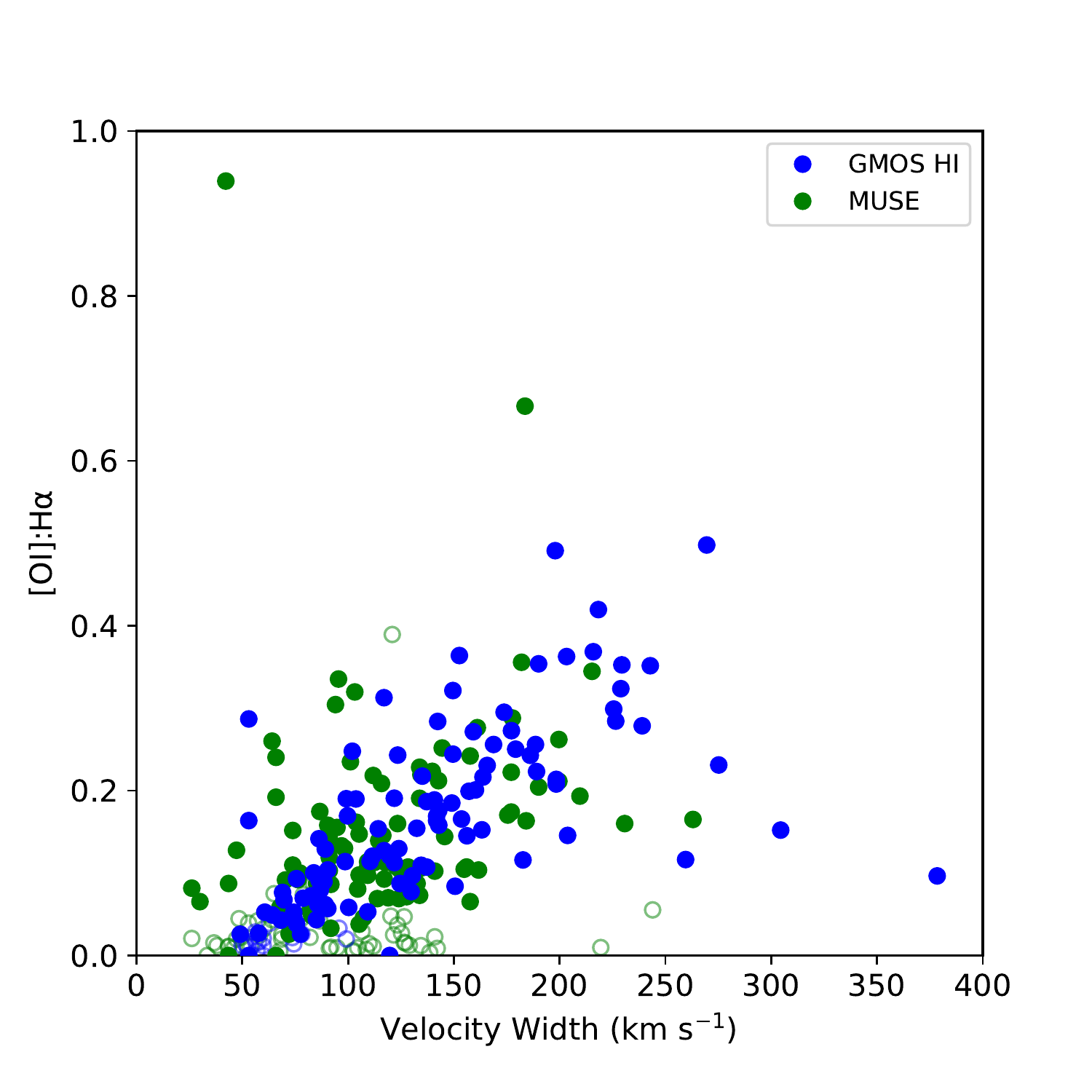}{0.33\textwidth}{(c)}}
\caption{(a) \sii\ density as a function of velocity FWHM.
(b) \sii:\ha\ ratio as a function of velocity FWHM.
(c) \oi:\ha\ ratio  as function of velocity FWHM.  As in Fig.~\ref{fig:correlations}, objects with \sii:\ha\ ratios $\ge$ or $<$ 0.4 are indicated by filled and open points, respectively.  
\label{fig:vel_trend}
 } 
\end{figure}

Furthermore, some of these shocks will be incomplete, in the sense that they will result from recent interactions of the SNR shock with dense cloudlets, and the shocks that are driven into these cloudets will  not have complete cooling and recombination zones; as a result emission lines from higher ionization potentials, such as \oiii, will be relatively brighter than those with lower ionization potentials, such as \ha\ and \oi. Nevertheless, and despite these concerns, it is useful to see the degree to which the observed line ratios are at least compatible with shock model expectations.

At present the most extensive set of published shock models of which we are aware are those that were presented by \cite{allen08} using the Mappings III code developed by \cite{sutherland93}.  The \cite{allen08} grids provide the expected fluxes for shocks for  velocities ranging from 100 to 1000 $\VEL$ for a range of pre-shock densities, magnetic field strengths and abundances.    The line fluxes presented as part of these models represent the total flux from such shocks, assuming that the shock is expanding into a uniform medium for a long enough time (and distance) that material behind the shock has cooled to a very low value behind the shock.  Such shocks are said to be ``complete''.   For the higher velocities, the time and distance involved can be quite large, and as a result the line fluxes and ratios are unlikely to be relevant for the optical lines of observed SNRs; hence we have limited our comparisons to velocities less than 400 $\VEL$.

Fig.\ \ref{fig:model_compare} provides a comparison between the observed line ratios we measure from SNR and SNR candidates in M83 to the line ratios calculated by \cite{allen08} for shocks in a medium with twice solar abundances, approximately that of M83 \citep{bresolin16}, and a pre-shock density of 1 cm$^{-3}$.  The various lines in the figure represent different pre-shock magnetic fields ranging from a negligible 10$^{-4}$ $\mu$G to 10 $\mu$G.  The line ratios shown are those calculated for the shock alone, and do not include emission from pre-shock gas.  
It is apparent that the observed line ratios do fall within the locus of the the various models that have been calculated.  The models show higher ratios of \sii:\ha, \nii:\ha\ and \oi:\ha\  as the velocities rise from  100 to 400 $\VEL$, which is qualitatively what we observe  based on the line broadening.   \cite{allen08} did not present results for similar twice-solar models with different pre-shock densities, but they did present models with  solar abundance for a variety of pre-shock densities; for densities 10$^{-1}$ to 10 cm$^{-3}$, the solar-abundance models look similar to those shown in Fig.\ \ref{fig:model_compare}.

As noted earlier, the velocity broadening we observe in the SNR spectra arises not from thermal broadening that occurs in the hotter plasma behind a shock but from the bulk motion of gas behind a shock.  The broadening we see reflects the combination of the shock velocity and the distribution of matter around the SNR, which is not uniform (as observed directly from images and indirectly in the asymmetric line profiles exhibited in some of the objects.). That said, we expect the observed velocity widths to be correlated with the typical shock velocities, in which case we expect there to be trends in the spectra with shock velocity.  \edit1{Among the spectral changes expected is a sharp increase in \ha\ (and other Balmer lines) with increasing shock velocity \citep{hartigan87}.}

As also shown in Fig.\ \ref{fig:vel_trend}, high \sii:\ha\ ratios and high \oi\ ratios are correlated positively with  velocity width, as expected.    \nii:\ha\ ratios are also correlated in this manner).  As also shown, the objects expanding into denser gas also tend to have higher velocity widths.  This can be understood if these are small diameter objects that are only detectable (see Sec. \ref{sec:galactic_trends} below) because they are expanding into denser material than is typical.

\subsection{Are Candidates that fail the \sii:\ha\ criterion SNRs? \label{sec:failed}}

We have argued that the objects observed to have  \sii:\ha\ greater than 0.4 are very likely SNRs.  There are, however, 72 objects with spectra in the catalog that fail the \sii:\ha\ $>$ 0.4  criterion.  These candidates, were, like the rest of the candidates in the catalog, originally identified on the basis of inspecting narrow-band interference-filter images. In these images, they appeared either as objects with elevated \sii:\ha\ ratios compared with nearby photoionized regions or in HST \feii\ 1.644\ $\mu$m images.  Of these, 22 show evidence of velocity broadening in their spectra, 21 have detectable \oi, 21 are positionally coincident with an X-ray source in the \cite{long14} {\em Chandra}
catalog, and 42 have associated \feii\ emission. Are these likely to be SNRs?  Some almost certainly are, if only due to uncertainties in the measured \sii:\ha\ line ratios, which as indicated in the comparisons made in Section \ref{sec:comparisons} is of order $\pm$0.07.  We will  conclude that those with \feii\ emission are almost certainly SNRs, since there are few if any mechanisms other than shocks that can produce \feii\ at the observed levels. However, many of the remaining objects must be considered marginal candidates at best.

The sources that fail the \sii:\ha\ criterion can be divided into two groups, the 28 candidates that fall within the nuclear region (GCD $<$ 0.5 kpc) and the 44 that reside in the outer galaxy.  The fraction of SNR candidates that fail the \sii:\ha\ criterion is not a strong function of source diameter, but at very small galactocentric distances, there are more sources that fail the  \sii:\ha\ criterion than that pass it.

The 28 sources in the nuclear region show a number of indications that they should be regarded as SNRs.  Most (22)  show evidence of line broadening; a substantial number (11) lie within 1\arcsec\ of a cataloged X-ray source;  six have detectable \oi\ emission; and 23 have associated \feii\ emission in HST data.  (Indeed, many of them were originally selected based on this criterion.)  Given that the nuclear region is  where background subtraction is most difficult and where there is likely to be additional variable mission along the line of sight, one might argue that the low  \sii:\ha\ line ratios arise from the confusion and background subtraction issues in this complex region. Indeed, some objects with higher derived ratios could just as easily be overestimated due to this effect.

On the other hand, if background-subtraction issues affect the line ratios, they almost certainly affect the reliability of the line broadening and the detection of \oi\ as well, especially for the line broadening, since the nuclear region is highly dynamic.  Given the number of X-ray sources in the nucleus, the fact that a number of the SNR candidates are spatially coincident with X-ray sources may not be surprising either.  Thus, the strongest evidence that these sources are actually SNRs is likely to be the associations of many with the detected \feii\ emission, which seems to  arise at detectable levels only from shocks \citep{blair14, long20}.

Of the 44 objects that fail the \sii:\ha$>0.4$ criterion and that lie outside of the nuclear region, none shows evidence of line broadening, 15 have \oi, and 10 are coincident with an X-ray source, all of which strengthen those objects as good SNR candidates. Some of the remaining objects may have low ratios due to  background subtraction issues that might affect individual objects (e.g. those in particularly complex extended emission regions), but some must be considered simply to be marginal SNR candidates with no other positive shock indicators.

{\bf Can [O~I]:\ha\ help?} As can be seen from Figures \ref{fig:model_compare} and \ref{fig:vel_trend} above, shock models predict and observations show that the  \oi:\ha\ ratio is elevated in shocks for much the same reason as \sii.  Recently, \citet{kopsacheili20}  have argued based on shock models presented by \citet{allen08} using the Mappings III code,  that in order to obtain a complete sample of SNRs in a galaxy, one should exploit the \oi:\ha\ ratio. More specifically, they suggest that \oi:\ha\ ratios as low as 0.01  still indicate evidence of shocks, and not photoionized gas.  They note that a number of the shock models in \citet{allen08} grids have \sii:\ha\ ratios less than 0.4, particularly in the velocity range 100-200 $\VEL$.   If they are correct, then a number of nebulae in our M83 sample with \sii:\ha\ ratios less than 0.4 should be considered SNRs.

However, we have concerns about adopting such a low \oi:\ha\ ratio criterion as proposed by \citet{kopsacheili20}.    First,  while \citet{kopsacheili20}  argued that \hii\ regions do not exhibit \oi:\ha\ ratios greater than 0.01, it is unclear that the models and observations for photoionized gas they used were representative of the low luminosity/surface brightness levels involved here, which approach that of the diffuse ionized gas (DIG).  Secondly, 
although we are aware of a few individual filament spectra in galactic SNRs that have \sii:\ha\ ratios somewhat below 0.4,, we are not aware of any global SNR spectra where the emission is characterized by a \sii:\ha\ ratio less than 0.4. Hence, low \sii:\ha\ objects with somewhat elevated \oi:\ha\ ratios must be considered questionable.

As to the DIG, \citet{reynolds98} used the WHAM instrument to measure \oi\ for three Milky Way DIG sight lines, seeing \oi:\ha\ in the range 0.012 – 0.044.  \citet{voges06} were able to observe faint \oi\ in M33 DIG by virtue of the blueshift of M33, finding ratios of 0.038, 0.040, and 0.097 for three regions in diffuse gas near the giant \hii\ region NGC~604.  \citet{rand98} observed the edge-on galaxy NGC~891 and was able to study the DIG \oi:\ha\ ratio with z-distance (which was effectively a study of \oi:\ha\ with surface brightness as well), finding ratios of 0.02-0.03 near the plane and values rising above 0.1 at the highest z values (lowest surface brightnesses) measurable.  Presumably, in viewing a face-on galaxy such as M83, spectra would be dominated by the (relatively) higher surface brightness DIG near the plane, with ratios below about 0.05.
 
In M83 we are aided by the observed $\sim$500 $\VEL$ redshift, which shifts the \oi\ $\lambda 6300$ line to the 6310-12 \AA\ region, making sky subtraction less of an issue. 
The observed \oi:\ha\ ratio trends downward with decreasing \sii:\ha\ and dips below 0.05 just as the \sii:\ha\ ratio decrease below 0.4.
Hence, the observed \oi:\ha\ ratios of these lower \sii:\ha\ nebulae are consistent with ratios seen in the DIG.  This is not to say that these nebulae are DIG, but rather that the \oi:\ha\ ratio cannot be used to discriminate shocks from DIG for these objects.

{\bf So exactly where does this leave us?}  While we are fairly confident that nearly all of the SNRs with \sii:\ha\ ratios greater than 0.4 are SNRs,  it is quite difficult to prove that the nebulae that have \sii:\ha\ ratios that are less than 0.4 are {\em not} SNRs, especially given the faintness of many of the nebulae.  Some of the objects that do not satisfy the \sii:\ha\ criterion are almost certainly SNRs, especially in the subsample of objects that were selected on the basis of \feii\ emission. Unfortunately, most of these are in the nuclear region,  where the observational uncertainties are largest.  We do not believe the detection of \oi\ has, as yet, proven to be determinative. That said, we encourage continued observations of all of  these nebulae and others like them, in the hope that ultimately  one can better understand how to obtain as complete a sample as possible of SNRs in M83 (and other galaxies). 
% \edit1 {In practice, it will probably never be possible to obtain a truly complete sample; rather, detectable SNRs will probably be limited by surface brightness.  We would not have been able to detect large, faint SNRs in M83 that are similar to the high-latitude ones in the Galaxy reported recently by \citet{fesen21}.  In estimating the limiting surface brightness for our sample, we refer not to the present spectroscopic observations, but to the sample measured with the MUSE IFU \citep{long20}.  There, the faintest SNRs have \ha\ surface brightness $\sim 3\times 10^{-17}$.  These are all ones located in the outer galaxy.  Closer to the galactic nucleus, where there is far brighter diffuse background, the limit may be as much as a factor of 10 brighter.}

\subsection{Variations of the properties of the SNRs in M83 \label{sec:galactic_trends}}

Ultimately, one would like to do more than simply catalog the SNRs in M83 and other nearby galaxies.   One would like to understand how the various properties of the SNRs reflect the nature of the SNR explosion and the environment which surrounded the SNR.   Some clues to this can be derived by looking at trends in the properties of the the SNRs as a function of the other known properties of a SNR, in this case properties like diameter and galactocentric distance.  Here we briefly reprise the more extensive discussion found in \cite{long22} based on the low resolution GMOS and MUSE data.
% Fig.8
\begin{figure}
\gridline{\fig{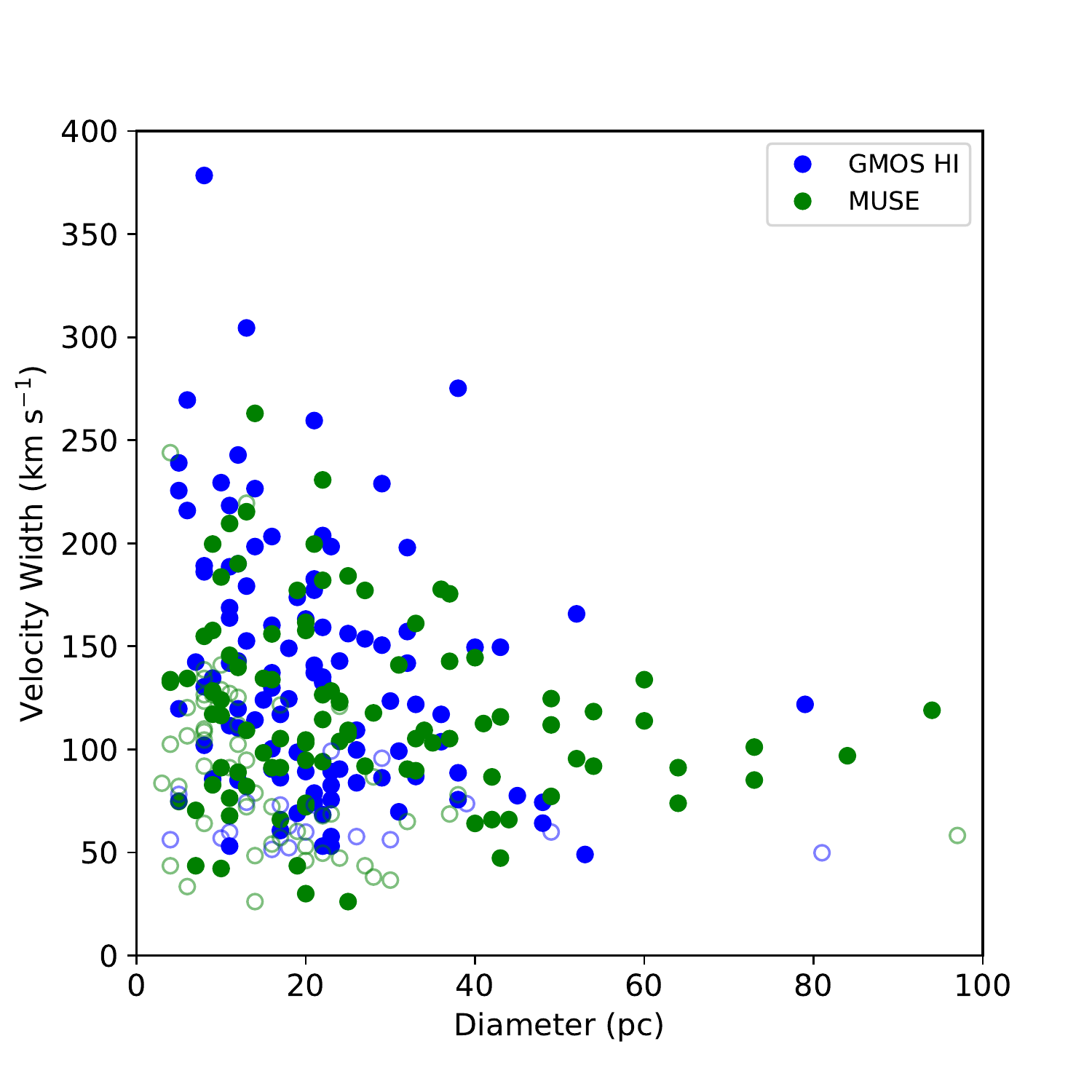}{0.33\textwidth}{(a)}
          \fig{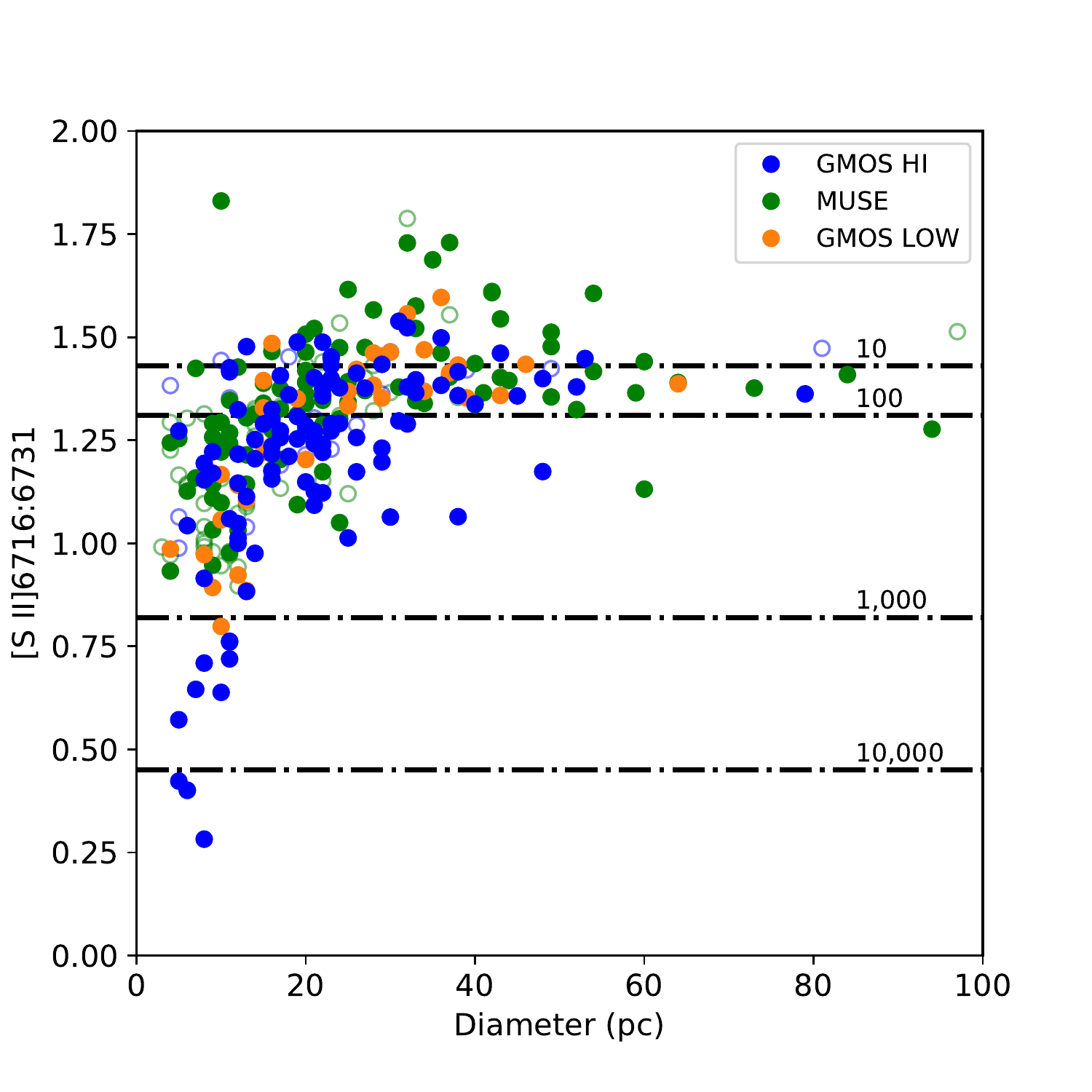}{0.33\textwidth}{(b)}
          \fig{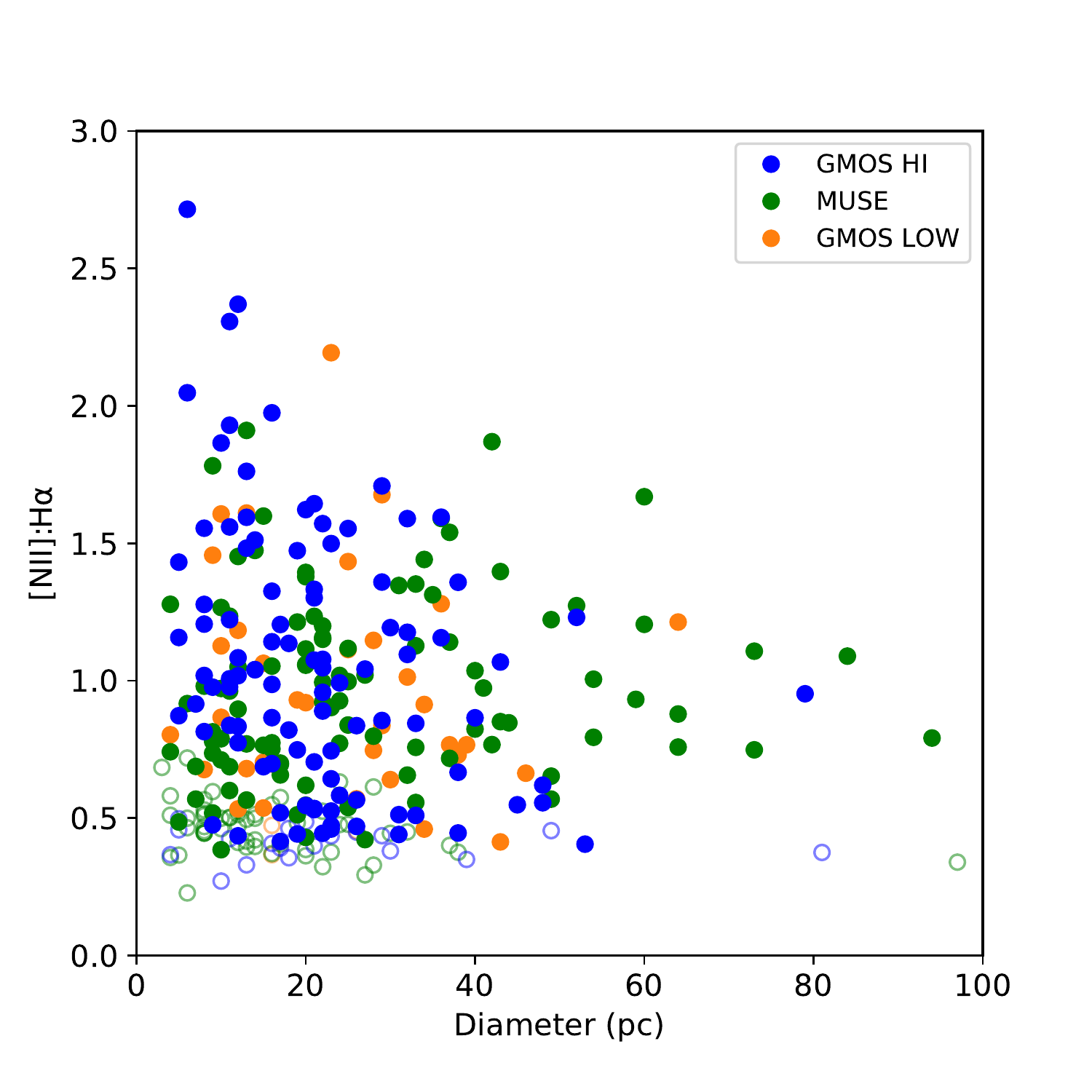}{0.33\textwidth}{(c)}}
\caption{(a) Velocity width as function of SNR diameter.
(b) Density-sensitive \sii\ ratio as a function of SNR diameter.  Broken lines indicate different density values.
(c) \nii:\ha\ ratio as a function of SNR diameter. In each of the plots, each SNR or SNR candidate appears only once, with precedence given to the observations taken with GMOS at high resolution, MUSE, and then GMOS at low resolution.  In all three panels, objects with \sii:\ha\ ratios $\ge$ 0.4 are plotted as solid markers, while those that fail this criterion are plotted as open markers in the relevant color.
\label{fig:D_trend} } 
\end{figure}

As indicated in Fig.\ \ref{fig:D_trend}, small-diameter objects show a much greater range of velocity widths compared to larger diameter ones, as one would expect if small-diameter objects are, on average, less evolved  than those at with larger diameters.  The range  of densities measured at small diameters is also larger, as one would expect if some objects are small not because they are young in an evolutionary sense, but because they are expanding into dense surrounding media.  We also see a larger range of \nii:\ha\ ratios at small diameters, which (from comparison with the \cite{allen08} models) could reflect a greater range of shock velocities in small-diameter SNRs.

Similar figures can also be made for the \oi:\ha\ and \sii:\ha\ ratio.  The \oi:\ha\ ratio vs. diameter plot is somewhat similar to the \nii:\ha\ in that the higher \oi:\ha\ ratio objects also tend to have diameters less than 30 pc; no trend is observed in the corresponding \sii:\ha\ diagram.

% Fig. 9
\begin{figure}
\gridline{\fig{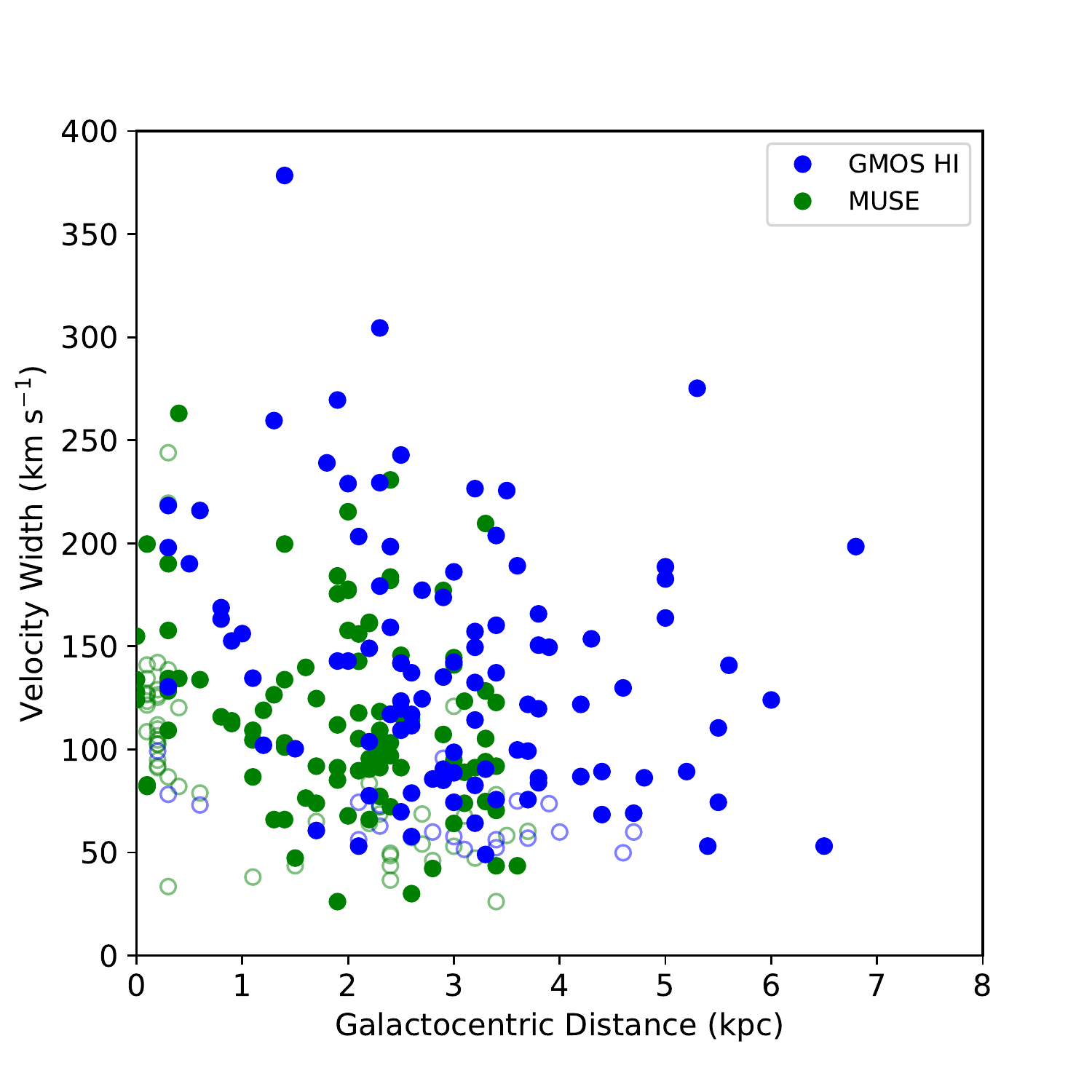}{0.33\textwidth}{(a)}
          \fig{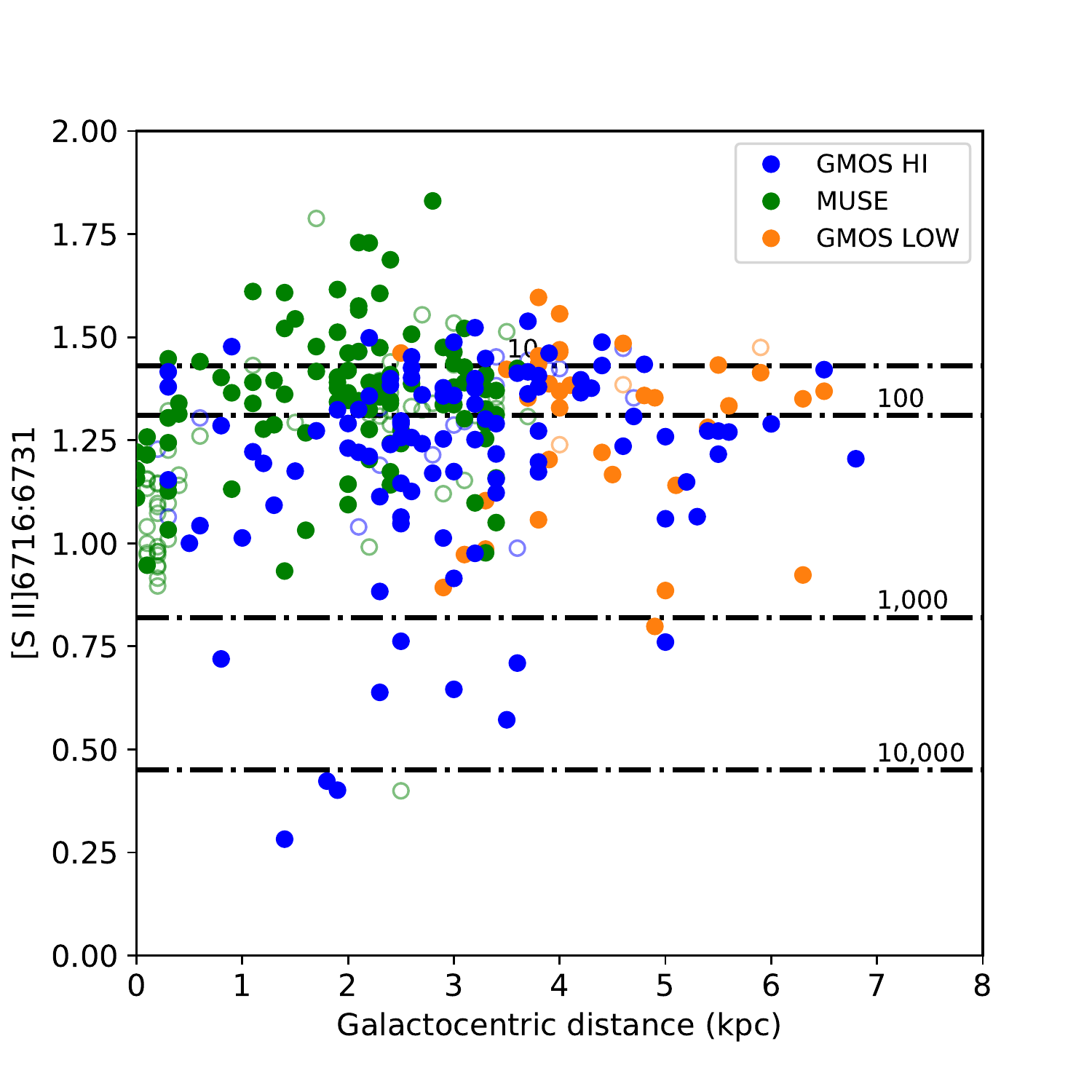}{0.33\textwidth}{(b)}
          \fig{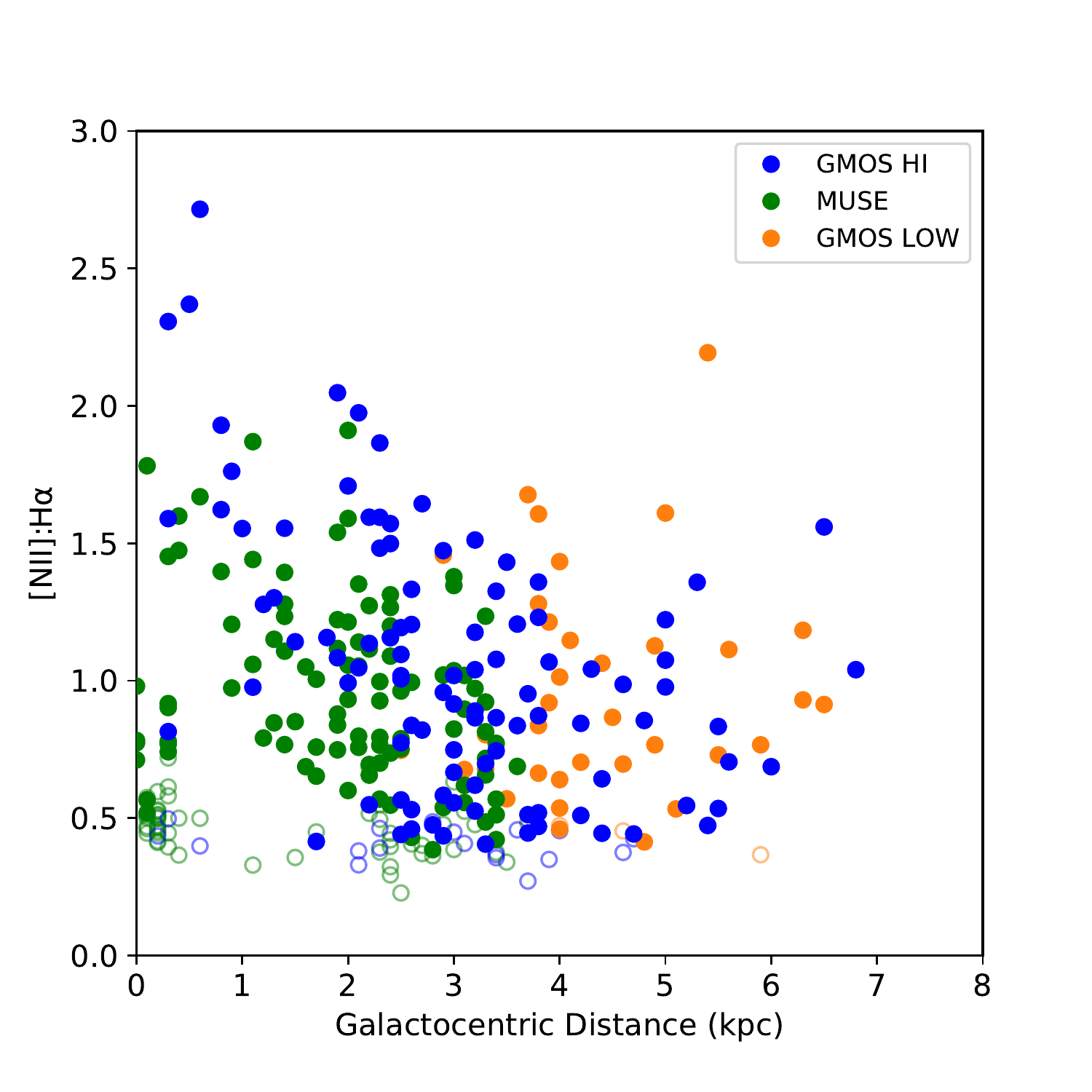}{0.33\textwidth}{(c)}}
\caption{(a) Velocity width as function of galactocentric distance.
(b) \sii\ density ratio as a function of galactocentric distance.
(c) \nii:\ha\ ratio as function of galactocentric distance.  As in Fig.~\ref{fig:D_trend}, objects with \sii:\ha\ ratios $\ge$ or $<$ 0.4 are indicated by filled and open points, respectively.
\label{fig:GC_trend}
 } 
\end{figure}

    As shown in Fig.\ \ref{fig:GC_trend}, there is no obvious trend of velocity width with galactocentric distance, except for the fact that more  of the SNRs (with \sii:\ha\  greater than  0.4) have detectable velocity broadening  at smaller than at larger galactocentric distances.  Outside the nucleus there is not much of a trend.  Similarly, except for the nuclear region there are no clear trends of \sii\ density.  There does appear to be a weak trend of the \nii:\ha\ ratio, that could be attributed to abundance gradients, but this is hard to disentangle from the trend that exists with SNR diameter.   Although various observers \citep[e.g.,][]{blair82,matonick97,galarza99,long18} have attempted to use SNR spectra to determine galactic abundance gradients, these data make clear that measurements using \hii\ regions are more reliable.

These results are not very different from those we reported in \cite{long22}, but the new GMOS data do increase the sample size and increase the number of SNRs observed at large galactocentric distances.  We interpret the trends to conclude that local conditions dominate.  With the exception of the \nii:\ha\ ratio the trends with diameter appear stronger than the trends with galactocentric radius.  SNRs encountering dense gas are brighter and detectable at small diameter than those encountering more tenuous media.  Large diameter SNRs are  detected only if the density of the ISM is relatively low, because if the density were high, a SNR would have already radiated away the kinetic energy of the SNR explosion and faded away.

\section{Summary and Conclusions}

We have reported high-resolution ($\sim 85\VEL$) spectroscopic observations of 119 SNRs and SNR candidates in M83 using the GMOS instrument on Gemini-South, including 24 objects for which there were no previously reported spectra.  The observations were designed to explore how velocity broadening could be used to provide an independent kinematic test of whether a SNR candidate was actually a SNR, compared to the more conventional test based on the  \sii:\ha\ ratio, and to explore how velocity broadening correlated with other properties of SNRs in M83.

Of the new sample observed here, we found 69 objects (58\%) that had lines broader than 100$\VEL$.  All of these objects had \sii:\ha\ ratios greater than 0.4.  None of the 19 SNR candidates that had observed \sii:\ha\ ratios less than 0.4 exhibited velocity broadening greater than 100$\VEL$.  Thus, the higher spectral resolution observations presented here provide independent confirmation of the SNR status of a significant number of SNR candidates, but have not helped significantly to elucidate the properties of the marginal ratio objects.  It remains possible that even higher spectral resolution could still shed light on the kinematics of these objects and whether they have lines significantly broader than those from \hii\ regions. However, higher resolution observations of a large set of such faint nebulae remain a significant observational challenge.

Based on the sample presented here, there is no indication that a survey based on high-resolution spectroscopy alone would be able to identify nebulae as SNRs that could not be found using \sii:\ha\ ratio criterion.  The only region where there were a significant number of candidate SNRs with \sii:\ha\ ratios less than 0.4, but line widths higher than typical \hii\ regions was in the nucleus of the galaxy, where a variety of factors (including the dynamic nature of the starburst region itself) could contribute to the apparently broader line widths observed.  

Comparing observed line profiles to Gaussian fits, it is clear that some objects have higher velocity, lower surface brightness components as well as asymmetries in their line profiles. This is to be expected since our spectra approximate global or nearly global spectra of each object. Both asymmetries and the broad wings can be understood as results of the SNR shocks encountering a complex surrounding multiphase ISM, giving rise to a range of shock velocities and shock brightnesses that contribute differently to the observed line profiles.  

In the M83 sample, the line ratios of \sii:\ha, \nii:\ha\ and \oi:\ha\ are correlated with one another, as is seen in shock models. SNRs that show significant velocity broadening tend to have the higher line ratios, as one would expect.  Small diameter objects tend to show shocks expanding into denser material than larger objects, as expected since these objects are brighter and then fade at smaller diameters than those expanding into low[density media which radiate the kinetic energy of an explosion over a longer period of time.

Finally, observations of M83 and/or other nearby galaxies at even higher spectral resolution than the $\sim 85 ~ \kms$ of those reported here will be useful for investigating some of the questionable objects.  Since photoionized nebulae should only exhibit motions of a few 10's of $\kms$ at most, and even older SNRs should have faster shocks that that, there is still parameter space where kinematics can be useful a diagnostic.  Unfortunately, such high resolution observations of faint extragalactic nebulae present a significant challenge to currently available instrumentation.

%\begin{acknowledgements}

\vspace{0.2in}
Based on observations obtained at the international Gemini Observatory, a program of NSF's NOIRLab, which is managed by the Association of Universities for Research in Astronomy (AURA) under a cooperative agreement with the National Science Foundation on behalf of the Gemini Observatory partnership: the National Science Foundation (United States), National Research Council (Canada), Agencia Nacional de Investigaci\'{o}n y Desarrollo (Chile), Ministerio de Ciencia, Tecnolog\'{i}a e Innovaci\'{o}n (Argentina), Minist\'{e}rio da Ci\^{e}ncia, Tecnologia, Inova\c{c}\~{o}es e Comunica\c{c}\~{o}es (Brazil), and Korea Astronomy and Space Science Institute (Republic of Korea).
We thank the staff at Gemini-South for their excellent support.
%In particular, we wish to thank Marcel Bergmann for assistance in refining routines for 1-D spectral extraction. 

We gratefully acknowledge several valuable suggestions by the referee, John Raymond.  PFW acknowledges financial support from the National Science Foundation through grant AST-0908566. WPB acknowledges support from the Dean of the Krieger School of Arts and Sciences and the Center for Astrophysical Sciences at JHU during this work.  

%\end{acknowledgements}

Facilities: Gemini:South (GMOS)

\bibliographystyle{aasjournal}

\bibliography{bibmaster}

\begin{thebibliography}{}
\expandafter\ifx\csname natexlab\endcsname\relax\def\natexlab#1{#1}\fi
\providecommand{\url}[1]{\href{#1}{#1}}
\providecommand{\dodoi}[1]{doi:~\href{http://doi.org/#1}{\nolinkurl{#1}}}
\providecommand{\doeprint}[1]{\href{http://ascl.net/#1}{\nolinkurl{http://ascl.net/#1}}}
\providecommand{\doarXiv}[1]{\href{https://arxiv.org/abs/#1}{\nolinkurl{https://arxiv.org/abs/#1}}}

\bibitem[{{Allen} {et~al.}(2008){Allen}, {Groves}, {Dopita}, {Sutherland}, \&
  {Kewley}}]{allen08}
{Allen}, M.~G., {Groves}, B.~A., {Dopita}, M.~A., {Sutherland}, R.~S., \&
  {Kewley}, L.~J. 2008, \apjs, 178, 20, \dodoi{10.1086/589652}

\bibitem[{{Blair} {et~al.}(1982){Blair}, {Kirshner}, \& {Chevalier}}]{blair82}
{Blair}, W.~P., {Kirshner}, R.~P., \& {Chevalier}, R.~A. 1982, \apj, 254, 50,
  \dodoi{10.1086/159703}

\bibitem[{Blair {et~al.}(2012)Blair, Winkler, \& Long}]{blair12}
Blair, W.~P., Winkler, P.~F., \& Long, K.~S. 2012, \apjs, 203, 8.
\newblock \url{http://stacks.iop.org/0067-0049/203/i=1/a=8}

\bibitem[{{Blair} {et~al.}(2014){Blair}, {Chandar}, {Dopita}, {Ghavamian},
  {Hammer}, {Kuntz}, {Long}, {Soria}, {Whitmore}, \& {Winkler}}]{blair14}
{Blair}, W.~P., {Chandar}, R., {Dopita}, M.~A., {et~al.} 2014, \apj, 788, 55,
  \dodoi{10.1088/0004-637X/788/1/55}

\bibitem[{{Blair} {et~al.}(2015){Blair}, {Winkler}, {Long}, {Whitmore}, {Kim},
  {Soria}, {Kuntz}, {Plucinsky}, {Dopita}, \& {Stockdale}}]{blair15}
{Blair}, W.~P., {Winkler}, P.~F., {Long}, K.~S., {et~al.} 2015, \apj, 800, 118,
  \dodoi{10.1088/0004-637X/800/2/118}

\bibitem[{{Bresolin} {et~al.}(2016){Bresolin}, {Kudritzki}, {Urbaneja},
  {Gieren}, {Ho}, \& {Pietrzy{\'n}ski}}]{bresolin16}
{Bresolin}, F., {Kudritzki}, R.-P., {Urbaneja}, M.~A., {et~al.} 2016, \apj,
  830, 64, \dodoi{10.3847/0004-637X/830/2/64}

\bibitem[{{Cox}(1972)}]{cox72}
{Cox}, D.~P. 1972, \apj, 178, 143, \dodoi{10.1086/151774}

\bibitem[{{Dopita} {et~al.}(2010){Dopita}, {Blair}, {Long}, {Mutchler},
  {Whitmore}, {Kuntz}, {Balick}, {Bond}, {Calzetti}, {Carollo}, {Disney},
  {Frogel}, {O'Connell}, {Hall}, {Holtzman}, {Kimble}, {MacKenty}, {McCarthy},
  {Paresce}, {Saha}, {Silk}, {Sirianni}, {Trauger}, {Walker}, {Windhorst}, \&
  {Young}}]{dopita10}
{Dopita}, M.~A., {Blair}, W.~P., {Long}, K.~S., {et~al.} 2010, \apj, 710, 964,
  \dodoi{10.1088/0004-637X/710/2/964}

\bibitem[{{Galarza} {et~al.}(1999){Galarza}, {Walterbos}, \&
  {Braun}}]{galarza99}
{Galarza}, V.~C., {Walterbos}, R.~A.~M., \& {Braun}, R. 1999, \aj, 118, 2775,
  \dodoi{10.1086/301113}

\bibitem[{{Hartigan} {et~al.}(1987){Hartigan}, {Raymond}, \&
  {Hartmann}}]{hartigan87}
{Hartigan}, P., {Raymond}, J., \& {Hartmann}, L. 1987, \apj, 316, 323,
  \dodoi{10.1086/165204}

\bibitem[{{Kopsacheili} {et~al.}(2020){Kopsacheili}, {Zezas}, \&
  {Leonidaki}}]{kopsacheili20}
{Kopsacheili}, M., {Zezas}, A., \& {Leonidaki}, I. 2020, \mnras, 491, 889,
  \dodoi{10.1093/mnras/stz2594}

\bibitem[{{Long} {et~al.}(1989){Long}, {Blair}, \& {Krzeminski}}]{long89}
{Long}, K.~S., {Blair}, W.~P., \& {Krzeminski}, W. 1989, \apjl, 340, L25,
  \dodoi{10.1086/185430}

\bibitem[{{Long} {et~al.}(1991){Long}, {Blair}, {Matsui}, \& {White}}]{long91}
{Long}, K.~S., {Blair}, W.~P., {Matsui}, Y., \& {White}, R.~L. 1991, \apj, 373,
  567, \dodoi{10.1086/170076}

\bibitem[{{Long} {et~al.}(2018){Long}, {Blair}, {Milisavljevic}, {Raymond}, \&
  {Winkler}}]{long18}
{Long}, K.~S., {Blair}, W.~P., {Milisavljevic}, D., {Raymond}, J.~C., \&
  {Winkler}, P.~F. 2018, \apj, 855, 140, \dodoi{10.3847/1538-4357/aaac7e}

\bibitem[{{Long} {et~al.}(2022){Long}, {Blair}, {Winkler}, {Della Bruna},
  {Adamo}, {McLeod}, \& {Amram}}]{long22}
{Long}, K.~S., {Blair}, W.~P., {Winkler}, P.~F., {et~al.} 2022, \apj, 929, 144,
  \dodoi{10.3847/1538-4357/ac5aa3}

\bibitem[{{Long} {et~al.}(2020){Long}, {Blair}, {Winkler}, \& {Lacey}}]{long20}
{Long}, K.~S., {Blair}, W.~P., {Winkler}, P.~F., \& {Lacey}, C.~K. 2020, \apj,
  899, 14, \dodoi{10.3847/1538-4357/aba2e9}

\bibitem[{{Long} {et~al.}(2014){Long}, {Kuntz}, {Blair}, {Godfrey},
  {Plucinsky}, {Soria}, {Stockdale}, \& {Winkler}}]{long14}
{Long}, K.~S., {Kuntz}, K.~D., {Blair}, W.~P., {et~al.} 2014, \apjs, 212, 21,
  \dodoi{10.1088/0067-0049/212/2/21}

\bibitem[{{Long} {et~al.}(2012){Long}, {Blair}, {Godfrey}, {Kuntz},
  {Plucinsky}, {Soria}, {Stockdale}, {Whitmore}, \& {Winkler}}]{long12}
{Long}, K.~S., {Blair}, W.~P., {Godfrey}, L.~E.~H., {et~al.} 2012, \apj, 756,
  18, \dodoi{10.1088/0004-637X/756/1/18}

\bibitem[{{Matonick} \& {Fesen}(1997)}]{matonick97}
{Matonick}, D.~M., \& {Fesen}, R.~A. 1997, \apjs, 112, 49,
  \dodoi{10.1086/313034}

\bibitem[{{McLeod} {et~al.}(2021){McLeod}, {Ali}, {Chevance}, {Della Bruna},
  {Schruba}, {Stevance}, {Adamo}, {Kruijssen}, {Longmore}, {Weisz}, \&
  {Zeidler}}]{Mcleod21}
{McLeod}, A.~F., {Ali}, A.~A., {Chevance}, M., {et~al.} 2021, \mnras, 508,
  5425, \dodoi{10.1093/mnras/stab2726}

\bibitem[{{Points} {et~al.}(2019){Points}, {Long}, {Winkler}, \&
  {Blair}}]{points19}
{Points}, S.~D., {Long}, K.~S., {Winkler}, P.~F., \& {Blair}, W.~P. 2019, \apj,
  887, 66, \dodoi{10.3847/1538-4357/ab4e98}

\bibitem[{{Rand}(1998)}]{rand98}
{Rand}, R.~J. 1998, \apj, 501, 137, \dodoi{10.1086/305814}

\bibitem[{{Reynolds} {et~al.}(1998){Reynolds}, {Hausen}, {Tufte}, \&
  {Haffner}}]{reynolds98}
{Reynolds}, R.~J., {Hausen}, N.~R., {Tufte}, S.~L., \& {Haffner}, L.~M. 1998,
  \apjl, 494, L99, \dodoi{10.1086/311154}

\bibitem[{{Saha} {et~al.}(2006){Saha}, {Thim}, {Tammann}, {Reindl}, \&
  {Sandage}}]{saha06}
{Saha}, A., {Thim}, F., {Tammann}, G.~A., {Reindl}, B., \& {Sandage}, A. 2006,
  \apjs, 165, 108, \dodoi{10.1086/503800}

\bibitem[{{Slavin} {et~al.}(2015){Slavin}, {Dwek}, \& {Jones}}]{slavin15}
{Slavin}, J.~D., {Dwek}, E., \& {Jones}, A.~P. 2015, \apj, 803, 7,
  \dodoi{10.1088/0004-637X/803/1/7}

\bibitem[{{Sutherland} \& {Dopita}(1993)}]{sutherland93}
{Sutherland}, R.~S., \& {Dopita}, M.~A. 1993, \apjs, 88, 253,
  \dodoi{10.1086/191823}

\bibitem[{{Turatto} {et~al.}(1989){Turatto}, {Cappellaro}, \&
  {Danziger}}]{turatto89}
{Turatto}, M., {Cappellaro}, E., \& {Danziger}, I.~J. 1989, The Messenger, 56,
  36

\bibitem[{{Voges} \& {Walterbos}(2006)}]{voges06}
{Voges}, E.~S., \& {Walterbos}, R.~A.~M. 2006, \apjl, 644, L29,
  \dodoi{10.1086/505575}

\bibitem[{{Williams} {et~al.}(2019){Williams}, {Hillis}, {Blair}, {Long},
  {Murphy}, {Dolphin}, {Khan}, \& {Dalcanton}}]{williams19}
{Williams}, B.~F., {Hillis}, T.~J., {Blair}, W.~P., {et~al.} 2019, \apj, 881,
  54, \dodoi{10.3847/1538-4357/ab2190}

\bibitem[{{Winkler} {et~al.}(2017){Winkler}, {Blair}, \& {Long}}]{winkler17}
{Winkler}, P.~F., {Blair}, W.~P., \& {Long}, K.~S. 2017, \apj, 839, 83,
  \dodoi{10.3847/1538-4357/aa683d}

\end{thebibliography}

\clearpage

%  SN1006 core sample paper,  Table 1
\begin{deluxetable}{lccccc}
%\rotate
%\tablewidth{600pt}
\tablewidth{0pt}
%\tabletypesize{small}
\tablecaption{Gemini-S/GMOS Multi-Object Spectroscopy Observations of M83 \label{tab:journal}}
\tablehead{
\colhead{Program} &
\colhead{Mask No.} &
\multicolumn{2}{c}{Center Position} &
\colhead{Date (UT)}  &
%\colhead{CWLs} &
\colhead{Total Exposure (s)\tablenotemark{a} }\\
\colhead{} & \colhead{} & \multicolumn{2}{c}{R.A.\phn\,(J2000.)\, Decl.} & \colhead{} & \colhead{}
%\colhead{Comments} 
%\colhead{$\rm \lambda_{c}[corr]$(\AA)\tablenotemark{b}} &
%\colhead{$\Delta \lambda$(\AA)\tablenotemark{a}} &
%\colhead {Exposure (s)} 
}
%\tabletypesize{\normal}
\tablewidth{0pt}
\startdata
%2 & 7 Apr 2011 & $7 \times 1000$ &  Variable clouds and seeing  \\[-8pt]
% &  & $7 \times 1000$ &    \\[-9pt]
 %& 7 Apr 2011 &  &  Variable clouds and seeing  \\[-9pt]
%2 &    &  $2 \times 1500$ &  \\[-1pt]
%&  9 Apr 2011  & $2 \times 1400$ & Clear; seeing $<1\arcsec$  \\[2pt]
%% & 9 Apr 2011 & $ 1500 + 1350 $ &  Clear; seeing $\lesssim 0\farcs 7$  \\
 GS-2019A-Q-303 & 1 & 13:37:08.5\phn\phn & \phn\phn-29:50:40 & 2019 May 8 & 2 CWLs\,$\times \,3 \times 1200$  \\[5pt]
 GS-2020A-Q-202 & 2 & 13:37:00.0\phn\phn & \phn\phn-29:52:48   & 2020 Feb 24-25 & 3 CWLs\,$\times \,3 \times 1200$ \\[5pt]
% & & & & 2 CWLs\,$\times 4 \times 1100$ \\[-7.5pt] 2 & 13:29:49.0\phn\phn & \phn\phn47:11:15   & 2020 Feb 24-25 & 3 CWLs\,$\times 3 \times 1200$ \\[2pt]
 GS-2021A-Q-235 & 3 & 13:36:52.5\phn\phn & \phn\phn-29:40:15   & 2021 Feb   10-14 & 3 CWLs\,$\times\, 3 \times 1200$ \\%[1pt]
  & 4 & 13:37:00.0\phn\phn & \phn\phn-29:52:48   & 2021 Jun 11 & 3 CWLs\,$\times\, 3 \times 1200$ \\%[1pt]
  & 5 & 13:37:08.5\phn\phn & \phn\phn-29:50:40   & 2021 Jun 12 & 3 CWLs\,$\times\, 3 \times 1200$ \\
\enddata
\tablenotetext{a}{Number of different Central Wavelength (CWL) Settings $\times$ number of exposures at each CWL $\times$ individual exposure time (s).}
\end{deluxetable}

% ksl - I do not understand why  \startlongtable was needed but th seemed to get the table referecing correct
\startlongtable
%Beginning of table
\begin{deluxetable}{crrr}
\tabletypesize{\scriptsize}
\decimals
\tablecaption{Comparison \hii\ Regions in M83}
\tablehead{
\colhead{Object} &
\colhead{RA} &
\colhead{Dec} &
\colhead{R}
\\
\colhead{} &
\colhead{(J2000)} &
\colhead{(J2000)} &
\colhead{(kpc)}
}
\startdata
HII-01 & 204.18766 & -29.83955 & 5.5 \\ 
HII-02 & 204.21082 & -29.88579 & 3.5 \\ 
HII-03 & 204.21168 & -29.84506 & 3.7 \\ 
HII-04 & 204.21303 & -29.84506 & 3.6 \\ 
HII-05 & 204.21485 & -29.90257 & 4.0 \\ 
HII-06 & 204.21510 & -29.88344 & 3.1 \\ 
HII-07 & 204.21549 & -29.90261 & 4.0 \\ 
HII-08 & 204.21603 & -29.90262 & 4.0 \\ 
HII-09 & 204.22645 & -29.84117 & 3.0 \\ 
HII-10 & 204.23437 & -29.82632 & 3.7 \\ 
HII-11 & 204.23665 & -29.79837 & 5.9 \\ 
HII-12 & 204.23695 & -29.88016 & 1.7 \\ 
HII-13 & 204.23813 & -29.91378 & 4.1 \\ 
HII-14 & 204.23886 & -29.84317 & 2.3 \\ 
HII-15 & 204.24208 & -29.88408 & 1.7 \\ 
HII-16 & 204.24918 & -29.86779 & 0.4 \\ 
HII-17 & 204.25016 & -29.87200 & 0.6 \\ 
HII-18 & 204.25450 & -29.85884 & 0.6 \\ 
HII-19 & 204.25625 & -29.85897 & 0.6 \\ 
HII-20 & 204.26000 & -29.86633 & 0.5 \\ 
HII-21 & 204.26361 & -29.82779 & 3.2 \\ 
HII-22 & 204.26399 & -29.85016 & 1.4 \\ 
HII-23 & 204.26579 & -29.90080 & 3.2 \\ 
HII-24 & 204.27361 & -29.83451 & 2.9 \\ 
HII-25 & 204.27728 & -29.84766 & 2.2 \\ 
HII-26 & 204.28000 & -29.88914 & 2.9 \\ 
HII-27 & 204.28254 & -29.88371 & 2.7 \\ 
HII-28 & 204.28518 & -29.84897 & 2.6 \\ 
HII-29 & 204.28654 & -29.85981 & 2.4 \\ 
HII-30 & 204.28990 & -29.85943 & 2.6 \\ 
HII-31 & 204.29062 & -29.87889 & 3.0 \\ 
HII-32 & 204.29147 & -29.85802 & 2.8 \\ 
HII-33 & 204.29237 & -29.85807 & 2.8 \\ 
HII-34 & 204.29303 & -29.85799 & 2.9 \\ 
HII-35 & 204.29522 & -29.85823 & 3.0 \\ 
HII-36 & 204.30774 & -29.88204 & 4.3 \\ 
\enddata
\label{h2_master}
\end{deluxetable}
% End of table

\startlongtable
%Beginning of table
\begin{deluxetable}{rrrrrrrrr}
\tabletypesize{\scriptsize}
\decimals
\tablecaption{M83 SNR and SNR candidate  spectra}
\tablehead{
\colhead{Source} &
\colhead{H$\alpha$ Flux$^a$} &
\colhead{[OI]6300$^b$} &
\colhead{H$\alpha$} &
\colhead{[NII]6584$^b$} &
\colhead{[SII]6716$^b$} &
\colhead{[SII]6731$^b$} &
\colhead{H$\alpha$ FWHM$^c$} &
\colhead{[S II] FWHM$^c$}
}
\startdata
B12-002 & 14.3$\pm$0.3 & 49.1$\pm$3.1 & 300 & 467.8$\pm$7.3 & 140.8$\pm$2.9 & 99.0$\pm$1.7 & 2.01$\pm$0.03 & 1.92$\pm$0.02 \\ 
B12-005 & 15.0$\pm$0.2 & 62.4$\pm$3.4 & 300 & 312.0$\pm$3.6 & 150.4$\pm$5.5 & 124.8$\pm$2.9 & 4.64$\pm$0.05 & 4.64$\pm$0.09 \\ 
B12-007 & 43.2$\pm$0.6 & 34.2$\pm$1.8 & 300 & 250.0$\pm$4.3 & 101.5$\pm$2.6 & 83.5$\pm$1.4 & 2.92$\pm$0.04 & 3.03$\pm$0.04 \\ 
B12-008 & 31.8$\pm$0.5 & 56.6$\pm$2.2 & 300 & 211.3$\pm$4.3 & 167.8$\pm$3.9 & 132.2$\pm$1.9 & 3.49$\pm$0.05 & 4.15$\pm$0.05 \\ 
B12-012 & 312.0$\pm$2.9 & 3.5$\pm$0.2 & 300 & 112.5$\pm$2.4 & 39.8$\pm$0.2 & 27.0$\pm$0.3 & 1.97$\pm$0.02 & 1.97$\pm$0.01 \\ 
B12-019 & 108.0$\pm$1.8 & 38.9$\pm$0.9 & 300 & 206.1$\pm$4.7 & 148.8$\pm$4.1 & 115.4$\pm$1.2 & 3.17$\pm$0.05 & 3.91$\pm$0.04 \\ 
B12-020 & 206.0$\pm$2.7 & 23.7$\pm$1.4 & 300 & 152.9$\pm$3.5 & 112.2$\pm$3.8 & 82.2$\pm$1.8 & 2.51$\pm$0.04 & 3.01$\pm$0.05 \\ 
B12-021 & 394.0$\pm$3.9 & 6.2$\pm$0.2 & 300 & 136.3$\pm$2.6 & 62.6$\pm$0.3 & 44.0$\pm$0.2 & 2.10$\pm$0.02 & 2.04$\pm$0.01 \\
\enddata
\tablenotetext{a}{Flux in units of 10$^{-17}$ ergs cm$^{-2}$ s$^{-1}$}
\tablenotetext{b}{Ratio to H$\alpha$  flux where, by convention, H$\alpha$ is normalized to 300.}
\tablenotetext{c}{In units of \AA.}
\label{snr_spectra}
\end{deluxetable}
% End of table

%Beginning of table
\begin{deluxetable}{rrrrrrrrr}
\tabletypesize{\scriptsize}
\decimals
\tablecaption{M83 \hii\ Region Spectra}
\tablehead{
\colhead{Source} &
\colhead{H$\alpha$ Flux$^a$} &
\colhead{[OI]6300$^b$} &
\colhead{H$\alpha$} &
\colhead{[NII]6584$^b$} &
\colhead{[SII]6716$^b$} &
\colhead{[SII]6731$^b$} &
\colhead{H$\alpha$ FWHM$^c$} &
\colhead{[S II] FWHM$^c$}
}
\startdata
HII-01 & 438.0$\pm$3.7 & 1.0$\pm$0.1 & 300 & 97.3$\pm$2.2 & 20.8$\pm$0.2 & 14.6$\pm$0.2 & 1.99$\pm$0.02 & 1.90$\pm$0.02 \\ 
HII-02 & 31.6$\pm$0.3 & 6.4$\pm$1.4 & 300 & 120.6$\pm$2.6 & 52.2$\pm$1.3 & 35.0$\pm$0.9 & 2.13$\pm$0.02 & 1.98$\pm$0.04 \\ 
HII-03 & 189.0$\pm$1.7 & 0.0$\pm$0.3 & 300 & 91.3$\pm$2.2 & 37.8$\pm$1.1 & 26.5$\pm$0.4 & 2.18$\pm$0.02 & 2.31$\pm$0.03 \\ 
HII-04 & 652.0$\pm$6.0 & 1.6$\pm$0.1 & 300 & 102.6$\pm$2.4 & 34.9$\pm$0.5 & 26.0$\pm$0.3 & 2.32$\pm$0.02 & 2.36$\pm$0.02 \\ 
HII-05 & 286.0$\pm$2.4 & 1.5$\pm$0.2 & 300 & 111.2$\pm$2.2 & 30.6$\pm$0.3 & 21.1$\pm$0.2 & 2.17$\pm$0.02 & 2.17$\pm$0.02 \\ 
HII-06 & 1110.0$\pm$10.5 & 2.2$\pm$0.3 & 300 & 112.2$\pm$2.4 & 34.0$\pm$0.3 & 23.9$\pm$0.3 & 1.97$\pm$0.02 & 1.91$\pm$0.02 \\ 
HII-07 & 799.0$\pm$7.0 & 1.1$\pm$0.1 & 300 & 111.5$\pm$2.2 & 24.9$\pm$0.2 & 17.8$\pm$0.2 & 2.16$\pm$0.02 & 2.15$\pm$0.02 \\ 
HII-08 & 253.0$\pm$2.5 & 1.4$\pm$0.3 & 300 & 119.8$\pm$2.6 & 29.3$\pm$0.3 & 20.5$\pm$0.2 & 2.03$\pm$0.02 & 2.02$\pm$0.02 \\
\enddata
\tablenotetext{a}{Flux in units of 10$^{-17}$ ergs cm$^{-2}$ s$^{-1}$}
\tablenotetext{b}{Ratio to H$\alpha$  flux where, by convention, H$\alpha$ is normalized to 300.}
\tablenotetext{c}{In units of \AA.}
\label{h2_spectra}
\end{deluxetable}
% End of table
             
\startlongtable
%Beginning of table
\begin{deluxetable}{lrrrrccccccl}
\tabletypesize{\scriptsize}
\decimals
\tablecaption{Properties of Supernova Remnants and Candidates in M83}
\tablehead{
\colhead{Source name$^a$} &
\colhead{RA} &
\colhead{Dec} &
\colhead{Diam.} &
\colhead{GCD} &
\colhead{X-ray$^b$} &
\colhead{\feii$^c$} &
\colhead{Spectra$^d$} &
\colhead{\sii:H$\alpha^e$} &
\colhead{\oi$^f$} &
\colhead{Vel.$^g$} &
\colhead{Comments}
\\
\colhead{} &
\colhead{(J2000)} &
\colhead{(J2000)} &
\colhead{(pc)} &
\colhead{(kpc)} &
\colhead{} &
\colhead{} &
\colhead{} &
\colhead{} &
\colhead{} &
\colhead{} &
\colhead{}
}
\startdata
B12-001 & 204.16663 & -29.85976 & 34 & 6.5 & -- & -- & GL & GL & GL & no & - \\ 
B12-002 & 204.16811 & -29.85181 & 11 & 6.5 & -- & -- & GH & GH & GH & no & - \\ 
B12-003 & 204.17040 & -29.85491 & 12 & 6.3 & X019 & -- & GL & GL & GL & no & - \\ 
B12-004 & 204.17292 & -29.87107 & 20 & 5.9 & -- & -- & none & no & no & no & - \\ 
B12-005 & 204.17325 & -29.83230 & 14 & 6.8 & -- & -- & GL-GH & GL-GH & GL-GH & GH & - \\ 
B12-006 & 204.17638 & -29.87149 & 24 & 5.7 & -- & -- & none & no & no & no & - \\ 
B12-007 & 204.17801 & -29.87637 & 12 & 5.5 & -- & -- & GH & GH & GH & GH & - \\ 
B12-008 & 204.18208 & -29.84608 & 21 & 5.6 & -- & -- & GH & GH & GH & GH & - \\
\enddata
\tablenotetext{a}{Source names beginning with B12, B14, D10 and L22 are from 
\cite{blair12}, \cite{blair14}, \cite{dopita10} and \cite{long22}, respectively.
}
\tablenotetext{b}{X-ray sources in the \cite{long14} Chandra catalog within 1\arcsec\ of an SNR candidate.}
\tablenotetext{c}{Detected or not in HST F164N [Fe II] imagery.  YES indicates sources candidates that were first identified as candidates based on [Fe II], while yes indicates that the source was detected in [Fe II], 
but identified by other means.  Sources that were in the region covered by the HST [Fe II] images but which were not detected are listed as no.  The remainder of the sources were outside the region covered in
the [Fe II images.}
\tablenotetext{d}{Source of spectra, if any, for the various SNRs and candidates.  
Objects labeled `GL' were observed by \cite{winkler17} using GMOS at low resoluation;
those labeled `M' were observed by \cite{long22} using MUSE; those labeled `GH' are 
from this work.
`All' implies that spectra were obtained in all 3 sets of observations.}
\tablenotetext{e}{Source of spectra that indicate a [S II]:H$\alpha$ ratio $>$0.4 for a particular object.}
\tablenotetext{f}{Source of spectra that  indicate an [O I]$\lambda$6300:H$\alpha$ ratio of greater than 3.3\% for a particular object.}
\tablenotetext{g}{Source of spectra that show deconvolved velocity broadening $>$100 $\rm km ~ s^{-1}$.}
\label{snr_summary}
\end{deluxetable}
% End of table

\begin{deluxetable}{lcccc}
%\tablewidth{0pt}
%\tabletypesize{small}
\tablecaption{Comparison of Criteria for SNRs in M83}
\tablehead{
\colhead{Criterion} &
\multicolumn{4}{c}{Number of Objects with Spectra\tablenotemark{a}}\\
\colhead{} &
\multicolumn{2}{c}{High-Res  Spectra Only  } &
\multicolumn{2}{c}{All Spectra }\\
\colhead{}  &
\colhead{Entire Galaxy} &
\colhead{Nucleus Excluded\tablenotemark{b}}  &
\colhead{Entire Galaxy} &
\colhead{Nucleus Excluded\tablenotemark{b}}\\
\colhead{}  &
\colhead{272 Total}  &
\colhead{225 Total}   &
\colhead{310 Total}  &
\colhead{263 Total}   
}
\tablewidth{0pt}
\startdata
  \sii:\ha $\,>0.4$		&	203   &   184  &  238  &    219  \\
  \oi:\ha $\,>0.03$		&	217   &   192   &  249  &   224  \\
 % (c) \feii\,$\lambda 1.64\, \mu$m detected\tablenotemark{c} & 122/234 & 86/200 & 133/256 & 96/209 \\
  $\Delta \upsilon > 100 \kms$\tablenotemark{c}   &	160   &   121   &  \nodata  & \nodata \\
 \sii:\ha $\,>0.4$ $and$  \oi:\ha $\,>0.03$  &      197   &  178  &  228   &  209   \\
 \sii:\ha $\,>0.4$ $and$ $\Delta \upsilon > 100 \kms$\tablenotemark{c}	     &      138   &  121  &  \nodata  &  \nodata \\
  \oi:\ha $\,>0.03$ $and$ $\Delta \upsilon > 100 \kms$\tablenotemark{c}	     &      142 &  121  &  \nodata &  \nodata \\
  \sii:\ha $\,>0.4$ $and$  \oi:\ha $\,>0.03$	$and$   $\Delta \upsilon > 100 \kms$\tablenotemark{c}      &      138   &  121  &  \nodata & \nodata \\
  \enddata
 
 \tablenotetext{a}{All numbers exclude the two young ejecta-dominated SNRs: SN1957d and B12-174a.}
 \tablenotetext{b}{Galactocentric radius $> $ 0.5 kpc.}
 \tablenotetext{c}{Included only for objects with high-resolution spectra; i.e., MUSE and GMOS High.}
\label{tab_criteria}
\end{deluxetable}

\end{document}